\documentclass[10pt]{article}
\usepackage{packages}
\usepackage{notations}
\usepackage{cite}
\usepackage{authblk}

\newcommand{\tikzOne}{
\begin{tikzpicture}[scale=1, every node/.style={scale=1, inner sep=0pt}]
\node[state,fill=black,minimum size=5pt]            (0) {~};
\node[state,right=6pt of 0,minimum size=5pt]            (1) {~};
\node[state,left=6pt of 0,minimum size=5pt]            (-1) {~};
\node[state,fill=black,left=6pt of -1,minimum size=5pt]            (-2) {~};
\node[state,fill=black,left=6pt of -2,minimum size=5pt]            (-3) {~};
\node[state,fill=black,left=6pt of -3,minimum size=5pt]            (-4) {~};
\node[state,left=6pt of -4,minimum size=5pt]            (-5) {~};
\node[state,left=6pt of -5,minimum size=5pt]            (-6) {~};

\node[state,fill=black,left=6pt of -6,minimum size=5pt]            (-7) {~};
\node[state,fill=black,left=6pt of -7,minimum size=5pt]            (-8) {~};
\node[state,fill=black,right=6pt of 1,minimum size=5pt]  (2) {~};
\node[state,fill=black, right=6pt of 2,minimum size=5pt]    (3) {~};

\node[state,right=20pt of 3,minimum size=5pt]                    (5) {~};
\node[state,right=6pt of 5,minimum size=5pt]                    (6) {~};
\node[state, right=6pt of 6,minimum size=5pt]                    (7) {~};
\node[state, right=6pt of 7,minimum size=5pt]                    (8) {~};
\node[state, right=6pt of 8,minimum size=5pt]                    (9) {~};
\node[state, right=6pt of 9,minimum size=5pt]                    (10) {~};
\node[right=30pt of 8]            (11) { \dots};

\node[left=30pt of 0]               (left-1) {~};
\node[left=100pt of 0]           (left-2) {\dots};
\node[left=40pt of left-2,minimum size=0pt]            (caption_a) {$\mathbf{(a)}$};
\node[right=31pt of 8]      (right-1) {~};

\node[above right= 10pt and 12pt of 3] (deli-1){~};
\node[below right= 12pt and 12pt of 3] (deli-2){~};

\node[rectangle,below delimiter=\}] (del-top-1) at ($0.5*(left-2.south) +0.5*(3.south)$) {\tikz{\path (left-2.south west) rectangle (3.north east);}};
\node[below=15pt] at ($0.5*(left-2.south) +0.5*(3.south)$.north) {$\mathcal{L}$};

\node[rectangle,above = 10pt, above delimiter=\{] (del-top-4) at ($0.5*(-5.south) +0.5*(0.south)$) {\tikz{\path (-7.south west) rectangle (-2.north east);}};
\node[above=30pt] at ($0.5*(-5.south) +0.5*(0.south)$.north) {$\ell_c$};

\node[rectangle,below delimiter=\}] (del-right-1) at ($0.5*(5.south) +0.5*(11.south)$) {\tikz{\path (5.south west) rectangle (11.north east);}};
\node[below=15pt] at ($0.5*(5.south) +0.5*(11.south)$.north) {$ \mathcal{R} $};

\draw[thick,dashed] (deli-1) --  (deli-2) ;
\node[below = 140pt of -3] (DensityMatrix){~};

\matrix [matrix of math nodes,left delimiter=(,right delimiter=),row sep=0.6cm,column sep=0.6cm, ampersand replacement=\&, above = 15pt of DensityMatrix] (m) {
~\&~\&~\&~ \\
~\&~\&~\&~ \\
~\&~\&~\&~ \\
~\&~\&~\&~ \\};
\draw[dashed] ($0.5*(m-1-2.north east)+0.5*(m-1-3.north west)$) --
      ($0.5*(m-4-2.south east)+0.5*(m-4-3.south west)$);
\draw[dashed] ($0.5*(m-2-1.south west)+0.5*(m-3-1.north west)$) --
      ($0.5*(m-2-4.south east)+0.5*(m-3-4.north east)$);
\node[above=-2pt of m-1-1] (top-1) {};
\node[above=-2pt of m-1-2] (top-2) {};
\node[above=-2pt of m-1-3] (top-3) {};
\node[above=-2pt of m-1-4] (top-4) {};

\node[below=5pt of $0.5*(m-1-1)+0.5*(m-1-2)$] (value-1) {$\hat{C}_0^{\mathcal{L}}$};
\node[below=4pt of $0.5*(m-3-1)+0.5*(m-3-2)$] (value-2) {$\mathbf{0}$};
\node[below=4pt of $0.5*(m-3-4)+0.5*(m-3-3)$] (value-3) {$\mathbf{0}$};
\node[below=5pt of $0.5*(m-1-4)+0.5*(m-1-3)$] (value-4) {$\mathbf{0}$};

\node[left=14pt of $0.5*(m-2-1.north west)+0.5*(m-3-1.north west)$] (left-5) {$\hat{C}_0=$};
\node[left=52pt of left-5,minimum size=0pt]            (caption_b) {$\mathbf{(b)}$};

\node[rectangle,above delimiter=\{] (del-top-1-2) at ($0.5*(top-1.north) +0.5*(top-2.north)$) {\tikz{\path (top-1.south west) rectangle (top-2.north east);}};
\node[above=12pt] at (del-top-1-2.north) {$\mathcal{L}$};

\node[rectangle,above delimiter=\{] (del-top-2-2) at ($0.5*(top-3.north) +0.5*(top-4.north)$) {\tikz{\path (top-3.south west) rectangle (top-4.north east);}};
\node[above=12pt] at (del-top-2-2.north) {$\mathcal{R}$};

\node[right = 150pt of $0.5*(m-2-4.north west)+0.5*(m-3-4.south west)$] (ReducedMatrix){~};
\node[right = 20pt of ReducedMatrix] (ReducedMatrix2){~};

\matrix [matrix of math nodes,left delimiter=(,right delimiter=),row sep=0.6cm,column sep=0.6cm, ampersand replacement=\&, left = 15pt of ReducedMatrix2] (mM) {
~\&~\&~\&~ \\
~\&~\&~\&~ \\
~\&~\&~\&~ \\
~\&~\&~\&~ \\};
\draw[thick, dashed] (mM-2-1.east) --(mM-4-3.east);
\draw[thick, dashed] (mM-1-2.north) --(mM-3-4.east);
\draw[thick] (mM-1-1) -- (mM-4-4);
\draw[>=open triangle 45, thin, <->] (mM-3-2.south east) -- (mM-2-3.north west);

\node[below=5pt of mM-3-2] (top-1-3) {$\ell_c$};

\node[left=10pt of $0.5*(mM-2-1.south west)+0.5*(mM-3-1.north west)$] (left-5-3) {$\hat{C}_0^{\mathcal{L}}=$};
\node[left = 30pt of left-5-3] (leftofCO){~};
\draw [->,>=stealth,shorten >=10pt,shorten <=4pt] (leftofCO.west) -- (left-5-3.west) ;

\end{tikzpicture}
}

\begin{document}
\title{Interplay between transport and quantum coherences in free fermionic systems}
\author[1]{Tony Jin}
\author[2]{Tristan Gauti\'e}
\author[3]{Alexandre Krajenbrink}
\author[1]{Paola Ruggiero}
\author[4]{Takato Yoshimura}

\affil[1]{\small DQMP, University of Geneva, 24 Quai Ernest-Ansermet, CH-1211 Geneva, Switzerland}
\affil[2]{\small LPENS, PSL University, CNRS, Sorbonne Universit\'e, Universit\'e de Paris, 24 rue Lhomond, 75231 Paris, France.}
\affil[3]{\small SISSA and INFN, via Bonomea 265, 34136 Trieste, Italy}
\affil[4]{\small Department of Physics, Tokyo Institute of Technology, Ookayama 2-12-1, Tokyo 152-8551, Japan}

\date{\today}

\maketitle

\begin{abstract}
We study the quench dynamics in free fermionic systems in the prototypical bipartitioning protocol obtained by joining two semi-infinite subsystems prepared in different states, aiming at understanding the interplay between quantum coherences in space in the initial state and transport properties. Our findings reveal that, under reasonable assumptions, the more correlated the initial state, the slower the transport is. Such statement is first discussed at qualitative level, and then made quantitative by introducing proper measures of correlations and transport ``speed''. Moreover, it is supported for fermions on a lattice by an exact solution starting from specific initial conditions, and in the continuous case by the explicit solution for a wider class of physically relevant initial states. In particular, for this class of states, we identify a  function, that we dub \emph{transition map}, which takes the value of the stationary current as input and gives the value of correlation as output, in a protocol-independent way. As an aside technical result, in the discrete case, we give an expression of the full counting statistics in terms of a continuous kernel for a general correlated domain wall initial state, thus extending the recent results in [Moriya, Nagao and Sasamoto, JSTAT 2019(6):063105] on the one-dimensional XX spin chain.
\end{abstract}

{\hypersetup{linkcolor=black}
\setcounter{tocdepth}{3}
\tableofcontents
}

\section{Introduction}~

How does quantum coherence affect a quantum many-body system far from equilibrium? Answering such a question not only enhances our understanding of the nature of dynamics in quantum systems, but also helps us design quantum materials with desired dynamical properties at will, something which is feasible today thanks to the unprecedented control in manipulating such systems~\cite{endres2016atom,ebadi2020quantum,scholl2020programmable,ManyBodySolitons}. Indeed, while initially quantum states in experiments have been prepared in either the ground state or in a thermal state of the system, today it is possible to engineer novel exotic states with peculiar coherent properties (e.g., quantum simulations \cite{Qsimulation1,Qsimulation2,Qsimulation3}, fragmented Bose-Einstein condensates \cite{evrard2020production,EntanglementBEC}). 
Such experimental realizations of correlated initial states in unconventional ways therefore call for a theoretical understanding of the effect of initial correlations on the subsequent dynamics.

Amongst possible situations where such attempts can be made, in the present work we focus on transport phenomena in one-dimensional systems.
The simple situation we have in mind is the extensively studied bipartitioning protocol~\cite{Schutz_XXzerotemps,PhysRevE.66.016135,PhysRevB.88.134301,Collura_2014,PhysRevB.89.214308,Viti_2016,Kormos_Ising, Gamayun_Freefermions,Perfetto_Ballisticfront}, where two semi-infinite systems are prepared in different states and then joined together. An ``imbalance'' between them will generate a flow from one side to another. 
In particular, we ask the following question: \textit{Does the presence of initial spatial correlations
have an effect on transport in quench protocols?}

For generic systems, hydrodynamic serves as a convenient tool to study transport problems, which has the advantage to be accessible also in presence of interactions. Hydrodynamics has been successfully applied to numerous systems, ranging from quark-gluon plasma \cite{Heinz2013} to graphene \cite{PhysRevLett.103.025301}, and more recently, to quantum integrable systems \cite{GHD_Ben_Olalla_Takato,PhysRevLett.117.207201}. In the latter context, it was termed generalized hydrodynamics (GHD), and has proved to be a powerful tool in characterizing the large-scale dynamics of integrable systems with a great analytic handle \cite{GHD_Ben_Olalla_Takato,PhysRevLett.117.207201,PhysRevLett.120.045301,SciPostPhys.3.6.039,PhysRevLett.121.160603}. Importantly, its validity has been confirmed in recent experiments \cite{PhysRevLett.122.090601,malvania2020generalized}. 

However, it is not clear whether this approach can give a complete answer to our question.
In fact, this theory is in general unable to capture effects due to correlations which are very far apart. One of its underlying assumptions, indeed, is that of local equilibration \emph{within fluid cells}, which means that different fluid cells are completely uncorrelated in this approximation, thus eventually leading to a partial loss of information with respect to the exact initial problem~\cite{Spohn1991}.
We mention, however, that there has been some (partially successful) attempts to extend GHD so to include such long-range correlations~\cite{PhysRevB.96.220302,10.21468/SciPostPhys.8.3.048,ruggiero2019conformal,ruggiero2020quantum}.

This motivates us to take a step back and consider the simplest example of integrable system, namely one-dimensional free fermions, for which numerous aspects can be studied analytically by means of \emph{exact} calculations.

\begin{figure}[t!]
\begin{centering}
\includegraphics[width=1.\textwidth]{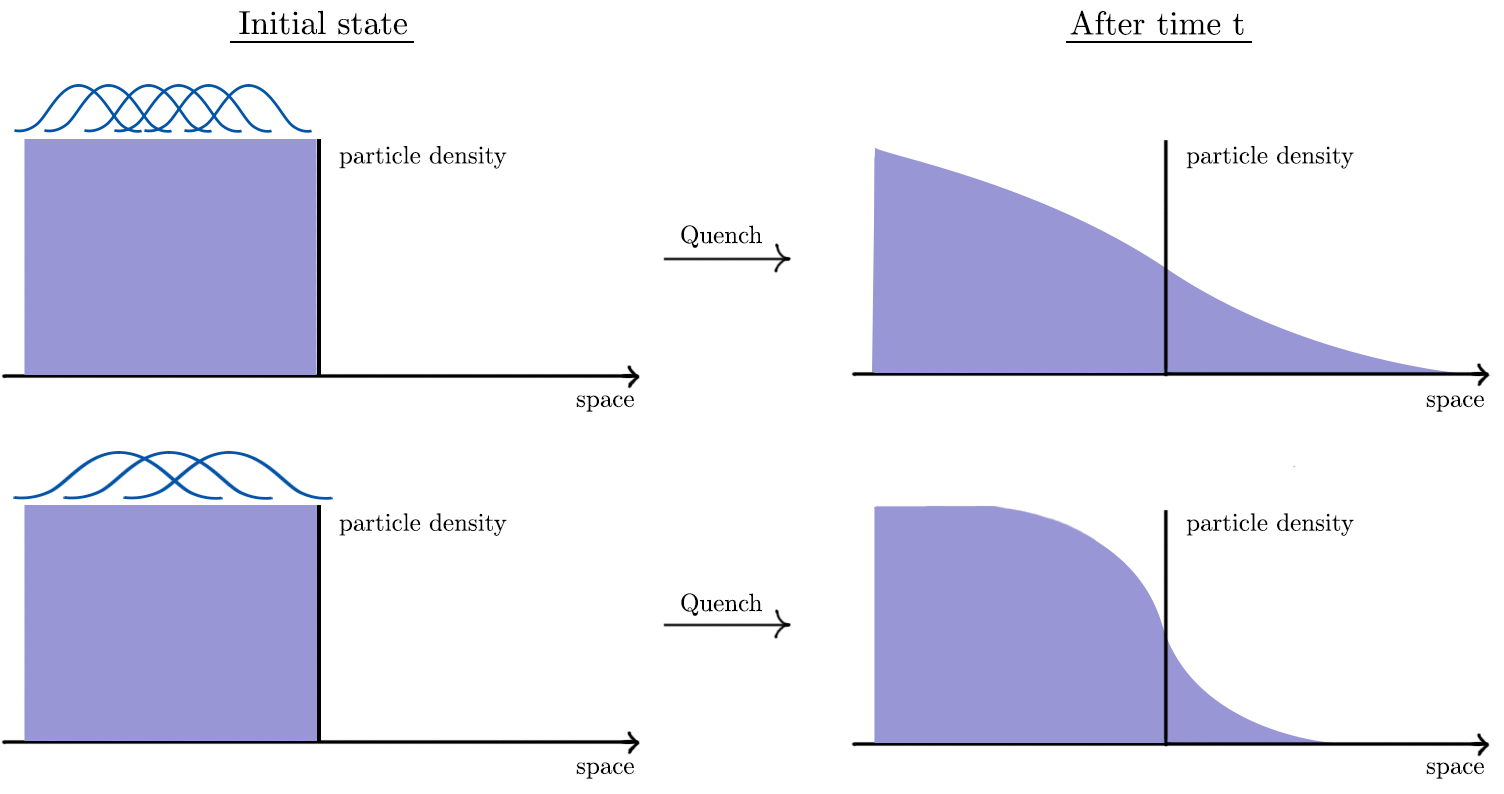}
\par\end{centering}
\caption{{\bf Left:} Initial state in the bipartitioning protocol where the left-half of the space is filled by particles with occupation $n_0$ and the right-half is empty. The top and bottom state are prepared in the same initial state but with different spatial correlation length.  {\bf Right:} Density profile after letting the system evolve for a time $t$. We will show that the more correlated the initial state, the greater the flow is at later times.}
\label{fig:quench}
\end{figure}

More specifically, we will consider the bipartite situation where the right ($\mathcal{R}$) side is initially empty, while the left ($\cal{L}$) side is prepared in a given state at the fixed value of the average density $ n_0$.
We will study how the flow is affected when one varies the typical length-scale over which particles in the $\cal{L}$-side are correlated (i.e., the \emph{coherence} length), as depicted in Fig~\ref{fig:quench}. 
Under physically plausible assumptions (that we are going to detail below, cf. Section~\ref{subsec:constraints}), we observe a clear phenomenon: \emph{the more correlated the state, the slower the transport}. To establish this statement at quantitative level, we shall introduce two quantities that measure the speed of (mass) transport and the strength of correlation, which will enable us to talk about these two different notions on an equal footing (Section~\ref{sec:quantitative}). 

Interestingly, this phenomenon is strongly linked to the integrable nature of the system, namely the existence of an infinite number of conserved quantities. A clear way to see that is the following. In the bipartition protocol, if we take two initial states with the same particle, momentum and energy densities, then conventional hydrodynamics~\cite{Spohn1991} (meaning for systems with few conservation laws, as opposite to the ``generalized'' one for integrable models~\cite{doyon2017large}) will predict the same flow for the two of them. However, for integrable models there may be an imbalance in the infinitely many other conserved charges of the system, giving rise to different currents ~\cite{doyon2017large}. As a striking illustration, it would be for instance possible to engineer two protocols I and II where the particle, momentum and energy densities are \emph{all higher} in protocol I but the flow would still \emph{be greater} in protocol II. This interplay between the flow of different conserved charges is another interesting observation of our analysis, and can be thought of as an analogue of the famous \emph{thermoelectric} effect \cite{Brantut2013} for integrable systems.

Finally, in relation to the initial issue of the limits of an hydrodynamic approach when describing long range correlations, this study further clarifies within the specific problem under investigation that, in the free case in continuous space at least, GHD can be though of as an \emph{exact} theory, rather then an approximation (in particular, no average over fluid cell is needed).
In fact our calculations are done in phase-space, relying on the so-called Wigner function~\cite{wigner,hillery1984distribution,cahill1969density} in its many-body version~\cite{bettelheim2006orthogonality,bettelheim2008quantum,bettelheim2011universal,bettelheim2012quantum,protopopov2013dynamics,dean2018wigner} (see Section~\ref{sec:ContinuousCase} for proper definitions), something which has been known and widely used~\cite{weinbub2018recent}.
On the other hand, in the free limit one can explicitly check that the GHD equation simply reduces to the (continuous) evolution equation for the aforementioned Wigner function. And since the latter can be derived without any further assumptions, the two approaches turn out to be equivalent, and, importantly, both exact.
This exact viewpoint of GHD is taken in Ref.~\cite{PhysRevB.96.220302,10.21468/SciPostPhys.8.3.048} for lattice free fermions, to derive the corrections due to the lattice in the form of Moyal expansion~\cite{moyal1949stochastic}. Further study is, instead, needed to understand the role of the hydrodynamic approximation (in particular, of fluid cells) in the interacting case. Some of the complications which occur in the attempt of extending our results in this direction are mentioned towards the end of the paper (Section~\ref{interacting}), but the general issue is left to future investigation.


Let us close this introduction by putting our specific quench problem in some more context within the existing literature on transport and correlations in free fermion, where in fact a great deal of analytic studies has been done in the past. 
Apart the standard references on bipartitioning protocols mentioned above, full counting statistics and large deviation in free fermionic systems have also been intensively studied \cite{Levitov1996,Schonhammer,Eisler_FCS,Yoshimura_2018,Sasamoto_LargedeviationXX,10.21468/SciPostPhys.8.3.036,Perfetto_Dynamics}.
The effect of quantum correlations in stochastic, diffusive free fermionic systems was investigated in quantum exclusion processes as well \cite{eisler2011crossover,ClosedQSSEP,OpenQSSEP,QKPZ}. 
The entanglement dynamics was considered both in homogeneous~\cite{fagotti2008evolution} and inhomogeneous~\cite{bertini2018entanglement} settings (and, remarkably, extended to interacting integrable models~\cite{alba2017entanglement,alba2018entanglement,alba2019quantum_2,alba2019towards,alba2019entanglement,mestyan2020molecular}).
More recently, the effect of continuous monitoring on the dynamics of a free fermionic chain was studied in \cite{TransportmonitoringBJS,10.21468/SciPostPhys.7.2.024}. 
Finally, transport properties in open free fermionic systems have been considered in \cite{maity2020growth,alba2021spreading,alba2021noninteracting,Generictransport}.

\paragraph{Outline.}The paper is structured as follows. We start in Section~\ref{sec:discreteCase} by considering a specific example of bipartite protocol for free fermions on a lattice. 
Through exact lattice calculation, we show explicitly that our main claim relating correlations and transport holds. 
An aside result of this Section is reported in Appendix~\ref{subsec:FCS}, specifically Eq.~\eqref{eq:ContinuousToDiscrete}, that expresses the full counting statistics in terms of a continuous kernel for a general correlated domain wall initial state, thus extending the recent results in \cite{Sasamoto_LargedeviationXX} on the one-dimensional XX spin chain.
In Section~\ref{sec:ContinuousCase}, we move to free fermions in the continuum where a great deal of simplifications occur. We can then consider a large family of initial states for which our statement still holds and, importantly, we manage to make it quantitative by introducing two quantities that measure the speed of (mass) transport and the strength of correlation, and further showing the existence of a \textit{transition map} 
which relate them in a protocol-independent way.
Finally, we comment on the more involved interacting case in Section~\ref{interacting}, and conclude by some perspectives and remarks in Section~\ref{conclusion}. Some derivations and technical details are reported in the remaining appendices. In particular, the full counting statistics for the half-filled protocol in the discrete case for arbitrary correlated initial conditions is presented there.

\section{An exact solution in the discrete case}
\label{sec:discreteCase}
We begin our investigation by studying a bipartitioning protocol in a free fermionic infinite one-dimensional lattice. 
In order to exemplify the relation between initial coherences and transport, we will consider a particular \emph{correlated domain-wall} (to be defined below), where the space is half-filled with fermions, tailored to have quantum coherence up to typical distance $\ell_{c}$, and study the effect on particles transport when varying $\ell_{c}$. 

The study of the uncorrelated domain-wall quench in the one-dimensional Heisenberg XX spin chain, equivalent to the free fermions on a lattice through a Jordan-Wigner transformation, has been the subject of many recent studies \cite{Schutz_XXzerotemps,Eisler_FCS,Sasamoto_LargedeviationXX}. These works focus, as we do, on a half-filled initial state, but with no coherence, corresponding to $\ell_{c}=1$ in our notation.

As an aside remark, we mention that while in the rest of the paper we will focus on the average particle number to characterize transport properties of the system, this is in general not sufficient for a full characterization. Instead, much more information can be extracted from higher momenta, all contained in the full counting statistics (FCS). While the latter is more difficult to study analytically, a result in this direction is reported in Appendix~\ref{subsec:FCS}, where we derive a Fredholm determinant representation of the FCS, when starting from an \emph{arbitrary} initial state with support on half of an infinite chain. This result extends the work~\cite{Sasamoto_LargedeviationXX} of Moriya, Nagao and Sasamoto, which was restricted to the uncorrelated case, to a generally correlated initial condition.
 
\subsection{The model and setup}
\label{subsec:generalSolXX}
 
The model under consideration is defined by the following Hamiltonian
\begin{equation} \label{Hdiscrete}
\hat{H} =- \frac{1}{2} \sum_{m=-\infty}^\infty \big( c_{m}^{\dagger}c_{m+1}+c_{m+1}^{\dagger}c_{m} \big) 
\end{equation}
where the creation and annihilation operators obey the fermionic algebra
\begin{equation}
\lbrace c_n^{\dagger}, c_m^{\dagger} \rbrace =\lbrace c_n, c_m\rbrace=0, \qquad  \lbrace c_n, c_m^{\dagger} \rbrace = \delta_{nm}, \qquad c_n ^2 = (c_n^\dagger)^2 = 0,
\end{equation}
with the anti-commutator defined by $\lbrace A,B \rbrace=AB+BA$. $\hat H$ can be diagonalized in Fourier space, with the fermionic momentum operators $\tilde{c}_k = \sum_{j=-\infty}^\infty e^{-\I j k } c_j$, and the single-particle spectrum reads $\{- \cos(k)\}_{k \in [-\pi,\pi]}$. 

We proceed by defining the time-dependent correlation matrix $\hat{C}(t)$ with matrix elements
\begin{equation}
{C}_{mn}(t)=\langle c_m^{\dagger}(t)c_n(t) \rangle 
\end{equation}
where $c_n(t) $ denote the time-evolved fermionic operators.
Standard calculations lead to the explicit and general expression
\be
\hat{C}(t) =\hat{J}(t)^\dagger{\hat{C}}_0 \hat{J}(t) \; .
\label{eq:solutiondiscrete}
\ee
where $\hat{C}_0 = \hat{C}(t=0)$ (which contains the information about the initial state) and $\hat{J}(t)$ has matrix elements ${J}_{mn}(t)=\I^{m-n}J_{m-n}( t)$ in terms of the $(m-n)^\mathrm{th}$ Bessel function of the first kind $J_{m-n} (t)$~\cite{NIST:DLMF}.

\subsection{A particular correlated domain-wall state}
\label{subsec:DiscreteProtocol}

We now specialize to a particular initial state. To this aim, we divide the lattice in two halves ${\cal L}:=\mathbb{Z}^{-}\cup \{ 0 \}$ and ${\cal R}:=\mathbb{Z}^{+}$. The $\cal{R}$-side is empty of particles, while the $\cal{L}$-side has fixed density $n_0$ and coherence length $\ell_c$, which is obtained by taking $n_{0}\ell_{c}$ particles living on $\ell_{c}$ sites as follows. 

First, for a given unit cell of $\ell_c$ sites, consider the so-called Dicke state \cite{Dickestate,bartschi2019deterministic}
\begin{equation} \label{Dicke-particular}
\ket{\phi^{\ell_{c}}_{n_0 \ell_c}} =\frac{1}{\sqrt{\binom{\ell_{c}}{n_{0}\ell_{c}}}}\sum_{{\cal C}(n_{0},\ell_{c})}\ket{{\cal C}} 
\end{equation}
where ${\cal C}(n_{0},\ell_{c})$ indexes all the possible ways of arranging $n_{0}\ell_{c}$ particles. This can be formulated as ${\cal C}(n_{0},\ell_{c})$ being a binary vector in $\{0,1 \}^{\ell_c}$ with $ n_0 \ell_c$ elements equal to $1$.
For instance, for $n_{0}\ell_{c}=2$ and $\ell_{c}=4$ : 
\begin{align*}
\ket{\phi^{\ell_{c}}} =\frac{1}{\sqrt{6}}\big(\ket{\bullet\bullet\circ\circ} +\ket{\bullet\circ\bullet\circ} +\ket{\bullet\circ\circ\bullet} +\ket{\circ\bullet\bullet\circ} +\ket{\circ\bullet\circ\bullet} +\ket{\circ\circ\bullet\bullet} \big).
\end{align*}
\begin{figure}[t!]
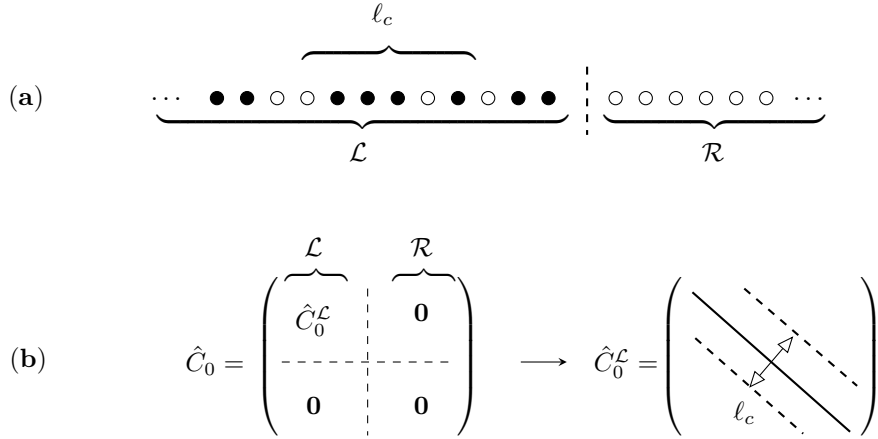

\begin{centering}
\tikzOne
\par\end{centering}
\caption{The initial state we consider in Section~\ref{sec:discreteCase}. \textbf{Panel (a):} Cartoon picture of the state. The lattice is half-filled with particle with mean density $n_{0}$ in the left-side ($\mathcal{L}$), while it is empty in the right-side ($\mathcal{R}$). In $\mathcal{L}$, particles are correlated on a typical distance
$\ell_{c}$ that we call the quantum coherence length. 
\textbf{Panel (b):}
Total two-point correlation function $\hat{C}_{0}^{\mathcal{L}}$ (right), and its restriction to the $\mathcal{L}$ sector, $\hat{C}_0^{\mathcal{L}}$, where matrix elements (given in \eqref{eq:discreteinitialcorrel}) decay to $0$ on a typical distance given by $\ell_{c}$.
Previous studies assumed $\hat{C}_{0}^{\mathcal{L}}$ to be completely diagonal.}
\label{fig:chain}
\end{figure}

Then, if $\rho^{\ell_{c}}:=\ket{\phi^{\ell_{c}}_{n_0 \ell_c}} \bra{\phi^{\ell_{c}}_{n_0 \ell_c}}$
is the associated density matrix, the total density matrix in the $\cal{L}$-side is taken as 
\begin{equation}
\rho_{{\rm tot}}=\frac{1}{\ell_{c}}\sum_{j=-(\ell_{c}-1)}^{0}\cdots\otimes\rho_{[-(2\ell_{c}-1+j),-(\ell_{c}+j)]}^{\ell_{c}}\otimes\rho_{[-(\ell_{c}-1+j),j]}^{\ell_{c}}
\end{equation}
where the brakets $[,]$ indicate the support of the density matrix.

The calculation of the corresponding two-point correlation function $ \hat{C}^{\rm tot}$ is straightforward, and its matrix elements read
\begin{align} 
C^{\rm tot}_{m, n}=\begin{cases}
n_{0} & \text{for }m=n\text{ and}\\
\frac{(\ell_{c}-|m-n|)}{\ell_{c}} \times c_{1,1+|m-n|}^{\ell_{c}} & \text{for }m\neq n\text{ and }|m-n|\leq\ell_{c}\\
0 & \text{otherwise}
\end{cases}
\label{eq:discreteinitialcorrel}
\end{align}
in terms of $c_{j,k}^{\ell_{c}}=\Tr(\rho^{\ell_{c}}c_{j}^{\dagger}c_{k})$, namely the two-point correlation within a coherent cell. Evaluating $c_{j,k}^{\ell_{c}}$
can be a tedious task because of the fermionic anti-commutation relations. This is nevertheless done, for completeness, in Appendix~\ref{subApp0:DetailsDiscreteProtocol} for the general case.
Here, instead, for simplicity, we restrict to the case where there is one particle in a given cell, i.e., $n_{0}={1}/{\ell_{c}}$, giving $c_{j,k}^{\ell_{c}}={1}/{\ell_{c}}$ for all
$ (j,k)$ in $[1,\ell_{c}]\times[1,\ell_{c}]$.
[For comparison, the uncorrelated domain-wall studied in~\cite{Schutz_XXzerotemps,Eisler_FCS,Sasamoto_LargedeviationXX} is instead represented by an initial diagonal correlation matrix with matrix elements $C_{m,n}^{{\cal D}}=n_{0}\delta_{m,n}$. Note that the two situations cannot be distinguished by only probing on-site observables.]

Finally, the total two point function is then $\hat{C}_0 = \hat{C}_0^{\mathcal{L}} \oplus  \hat{C}_0^{\mathcal{R}}$, with $\hat{C}^{\mathcal{R}}_0 = \hat{C}^{\rm tot}$ and $\hat{C}^{\mathcal{R}}_0 =0$, as shown in Fig.~\ref{fig:chain}. Such initial condition together with Eq.~\eqref{eq:solutiondiscrete}, gives the local density $C_{m,m}(t)$ at site $m$ as (see Appendix~\ref{subApp2:DetailsDiscreteProtocol} for details).

\begin{align}
C_{m,m}(t) & =n_{0}B_{t}(m,m)+2\sum_{k=1}^{\ell_{c}}\cos(\frac{k\pi}{2})C_{k}B_{t}(m+k,m)
\label{eq:solutionCdiscrete}
\end{align}
where $B_{t}(m,n)$ is the discrete Bessel kernel (cf. Eq.~\eqref{eq:BesselDisKer} in Appendix~\ref{subApp2:DetailsDiscreteProtocol}) and $C_{j_1-j_2}:=(\hat{\mathcal{C}}_0^{\mathcal{L}})_{j_1,j_2}(t=0)$. 
We define the \emph{hydrodynamic limit} $\lim_{\rm{h}}$ by the limit where $t\to \infty$, $m \to \infty$ while $u:=\frac{m}{t}$ is kept fixed. Let $\Phi(u)=\lim_{\rm{h}} C_{m,m}(t)$, we then have (see Appendix~\ref{subApp3:DetailsDiscreteProtocol} for the proof)

\begin{figure}[t!]
    \centering
    \includegraphics[scale=0.6]{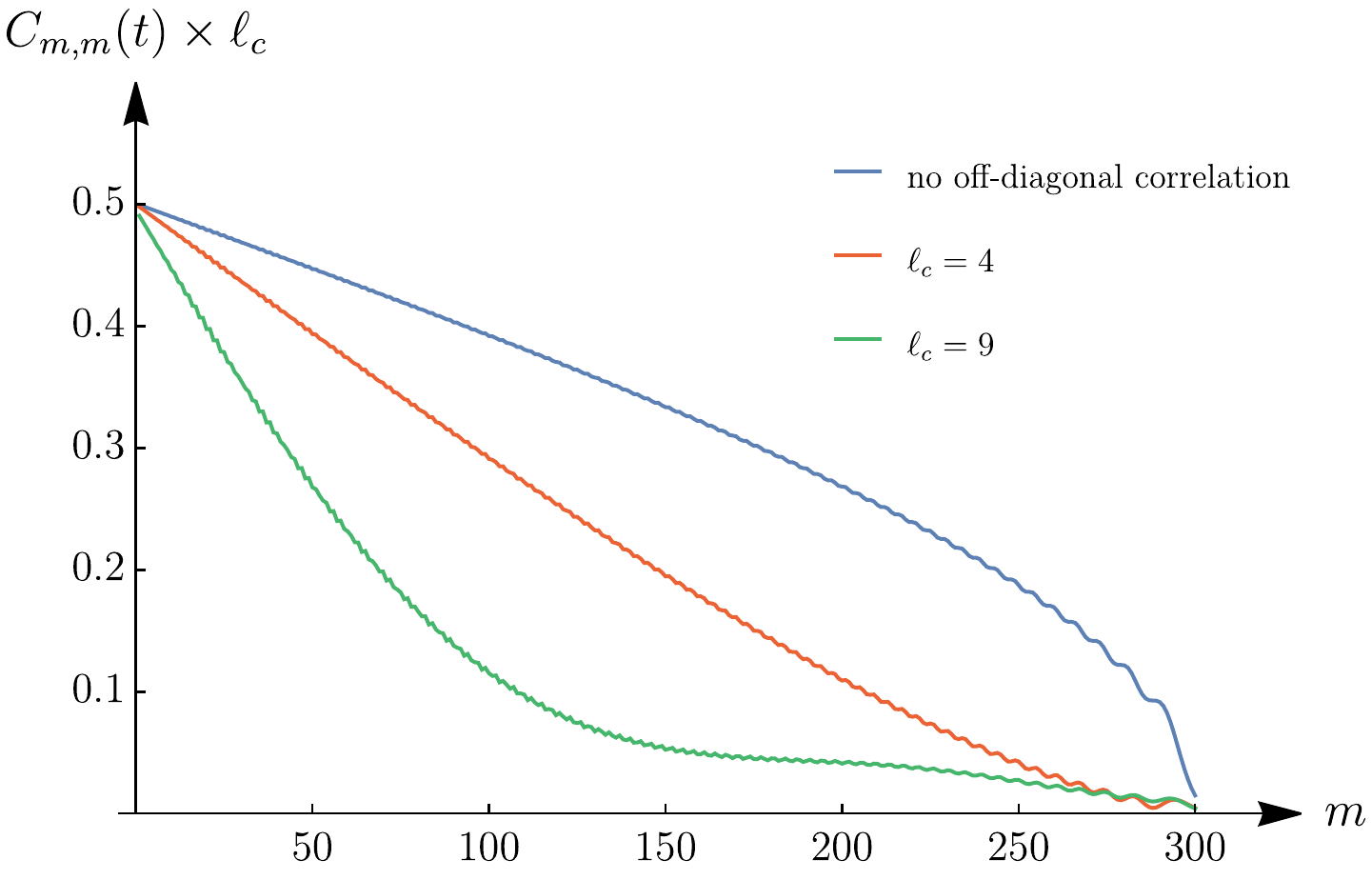}
\caption{Density profile at $t=300$ in the right half-space for difference
coherence length. For readability, all the curves have been rescaled
by a factor $\ell_{c}$ to match at the origin.}
\label{fig:discrete_coherence}
\end{figure}
\begin{equation}
    \Phi(u)= \frac{\arccos(u)}{\pi \ell_c}+
    \frac{2}{\pi }  \sum_{k=1}^{\ell_c}\cos\left( \frac{k\pi}{2} \right) \frac{\ell_c-k}{k \ell_c^2} \sin(k \arccos(u))
    \label{eq:krajDensityLargeTime}
\end{equation}
In Fig.~\ref{fig:discrete_coherence} we plotted for fixed time $t=300$
the density profile in the $\cal{R}$-part of the system, for different choice of $\ell_{c}$. For readability, all the
curves are rescaled so that they match at the origin. We clearly see
that the transport is affected by the coherence present in the initial
state. Specifically, that \emph{the more correlated} the state, \emph{the slower} is
the transport. 

\begin{figure}[t!]
\centering
\includegraphics[scale=0.6]{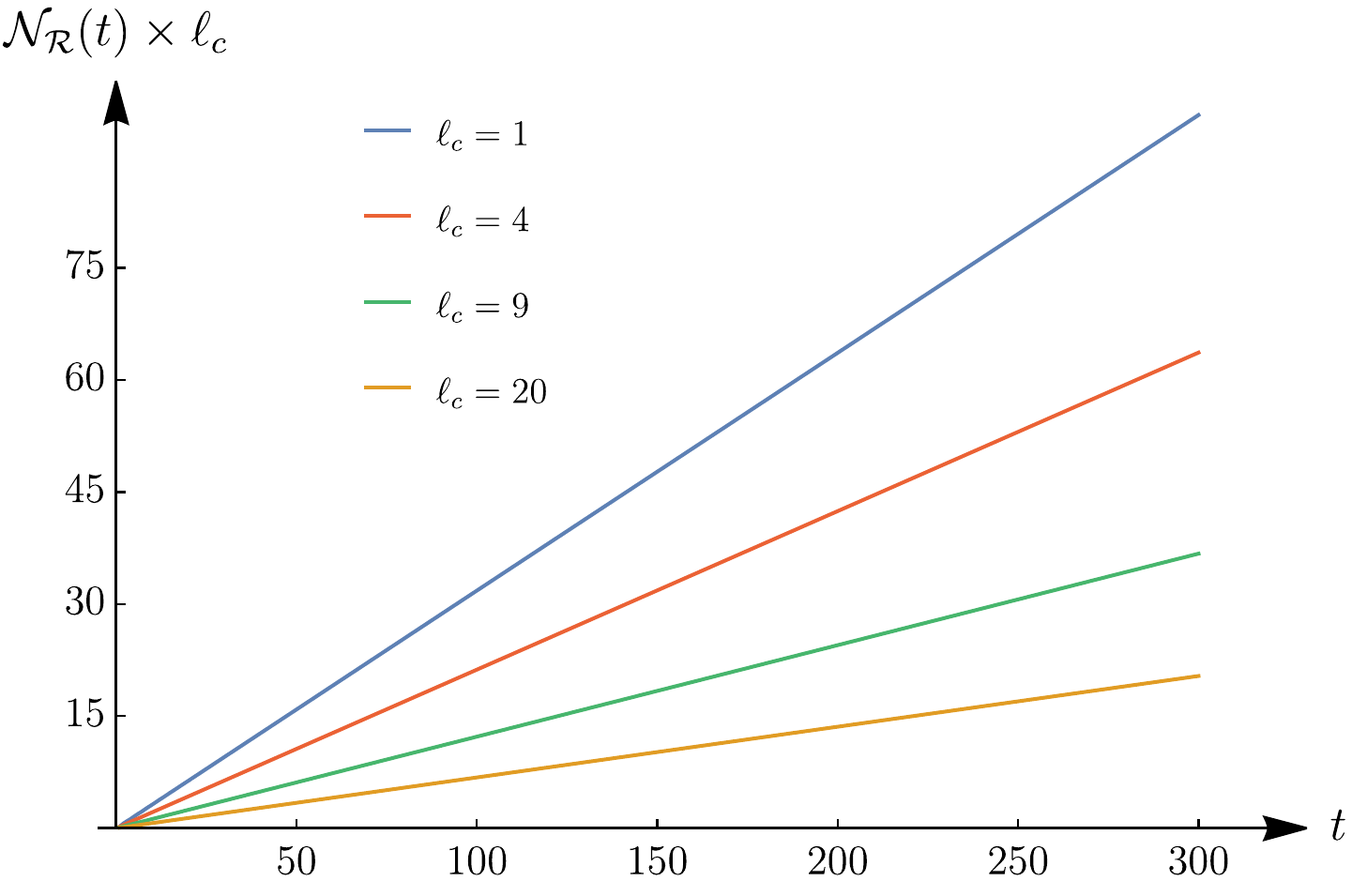}
\caption{Total number of transferred particles $\mathcal{N}_{{\cal R}}(t)$ as a function
of time for different coherence length. All the curves have
been rescaled by $1/\ell_{c}$ for readability. We see that
$\mathcal{N}_{{\cal R}}(t)$ scales linearly with time
and that the slope decreases as the coherence length $\ell_{c}$ increases. }
\label{fig:TotalParticleNumber}
\end{figure}

From the local density, one can further obtain the total number of particle in ${\cal R}$ at time $t$ as 
\begin{equation} \label{N_R}
    \mathcal{N}_{{\cal R}}(t)=\sum_{m\in \cal R} C_{m,m}(t). 
\end{equation}
This quantity measures the number of fermions which did flow from the left half of the lattice to the right one.
For a time $t$ larger than the characteristic length $\ell_c$, the transport is ballistic and ${\cal N}_{{\cal R}}(t)$
scales linearly with time 
\begin{equation}
    {\cal N}_{\cal R}(t) \simeq \alpha_T t,
\end{equation}
with (see Appendix~\ref{subApp3:DetailsDiscreteProtocol})
\begin{equation}
\ell_c \alpha_T=\frac{1}{\pi}- \frac{2}{\pi \ell_c}  \sum_{k=2}^{\ell_c}\cos\big(\frac{k\pi}{2}\big)^2 \frac{\ell_c-k}{k^2-1}.
\label{eq : discreteslope}
\end{equation}
From this expression, one sees that increasing $\ell_{c}$ slows the transport down. This is also clearly depicted in Fig.~\ref{fig:TotalParticleNumber}. 


\section{A general framework in the continuous case }
\label{sec:ContinuousCase}

We now turn to the analysis of the problem in the continuous limit, aiming at understanding the effects of initial correlations on {transport}. 
We consider again a bipartitioning protocol where
half-space ${\cal{L}} := (-\infty,  0  ] $ is filled and the rest ${\cal{R}} := [ 0 ,\infty)$ is empty, and study various protocols, where the $\cal{L}$-part has always the same fixed density $n_0$, but the coherence properties are different. 

\subsection{The model}
\label{subsec:ContinuousModel}
Let $\Psi^{\dagger}(x),\Psi(x)$ be continuous fermionic fields satisying the anticommutation relation $\{\Psi(x),\Psi^{\dagger}(x')\}=\delta(x-x')$, and associated to the free Hamiltonian
\begin{equation} \label{Hcontinuum}
H=\int_\mathbb{R} \rmd x\frac{\hbar^{2}}{2m}\partial_{x}\Psi^{\dagger}\partial_{x}\Psi,
\end{equation}
where $m$ is the mass of the system.

A central object to our study is the \emph{Wigner function}\cite{wigner} defined as
\begin{align}
n(x,k ,t) & :=\int_\mathbb{R} \rmd y e^{\I ky}\Tr(\hat{\rho}_{t}\Psi^{\dagger}(x+y/2)\Psi(x-y/2))\\
 & =\int_\mathbb{R} \rmd y e^{\I ky}C(x-y/2,x+y/2, t).
\end{align}
When integrated over space (respectively, over momenta), the Wigner function gives the fermion
occupation number in the momentum basis (respectively, in the space basis). 
Moreover, there is a simple inverse transform allowing to retrieve the two-point correlation
function, $C(x,y,t):=\Tr(\hat{\rho}_t\Psi^{\dagger} (x)\Psi (y))$ (here $\hat{\rho}_t$ is the density matrix at time $t$), as
\begin{equation}
C(x,y,t)=\int_\mathbb{R}\frac{\rmd k}{2\pi}n \left(\frac{x+y}{2},k,t\right)e^{\I k(x-y)}\label{eq:TFinverse}
\end{equation}
Hence, obtaining the Wigner function amounts to solve the problem at the level of two-point correlation function. For translationally invariant states, we further have $C(x,y,t)=C(x-y,0,t)$, so that the Wigner function is solely the Fourier transform $\mathcal{F}$ of the two-point correlation function. In this case, as a shorthand, we will denote it as $C(r,t)$ (with $r=x-y$). 

As mentioned in the introduction, the Wigner function is nothing but the free version of the occupation number function for GHD \cite{GHD_Ben_Olalla_Takato}. 
In fact, the equation governing its evolution can be deduced directly from the infinitely many continuity equations associated to the likewise infinite local conserved charges ${q_i(x,t)}$ the model possesses. To see this, let $h_i(k)$ be the one-particle eigenvalue associated to the charge $\hat{Q}_i=\int_\mathbb{R} \rmd x q_i (x,t)$, we then have~\cite{fagotti2016charges} $\langle q_i(x,t)\rangle=\int_\mathbb{R} \frac{\rmd k}{2\pi} n(x,k,t) h_i(k)$, and $\langle j_i(x,t)\rangle=\int_\mathbb{R} \frac{\rmd k}{2\pi} \frac{\hbar k}{m}n(x,k,t) h_i(k)$, where $\langle\bullet\rangle:=\mathrm{Tr}(\varrho\,\bullet)/\mathrm{Tr}\,\varrho$ and $j_i(x,t)$ is the current operator associated to $\hat{Q}_i$. Importantly these expressions are valid for any $x$ and $t$, which allows us to plug them into the continuity equation $\partial_t\langle q_i(x,t)\rangle+\partial_x\langle j_i(x,t)\rangle=0$. Invoking that $\{h_i(k)\}$ form a complete basis of the space $L^2(\mathbb{R},n(k)/2\pi)$, it follows that the dynamical evolution for $n(x,k,t)$ generated by the hamiltonian \eqref{Hcontinuum} is given by the transport equation
\begin{equation}
\partial_{t}n(x,k,t)+\frac{\hbar k}{m}\partial_{x}n(x,k,t)=0,
\label{eq:continuityequation}
\end{equation}
This is simply solved by
\begin{equation}
n(x,k,t)=n\left(x-\frac{\hbar k}{m}t,k,0 \right).
\end{equation}
The physical intuition about the evolution equation~\eqref{eq:continuityequation} is straightforward. The different propagating modes, indexed continuously by momentum $k$, are independent from one another. 
We stress that we do not need to interpret $n(x,k,t)$ as a coarse grained slowly varying distribution function, as one would do in a hydrodynamic theory. Conversely, this is an exact microscopic quantity.

\begin{figure}[t!]
\centering
\includegraphics[width=0.5\textwidth]{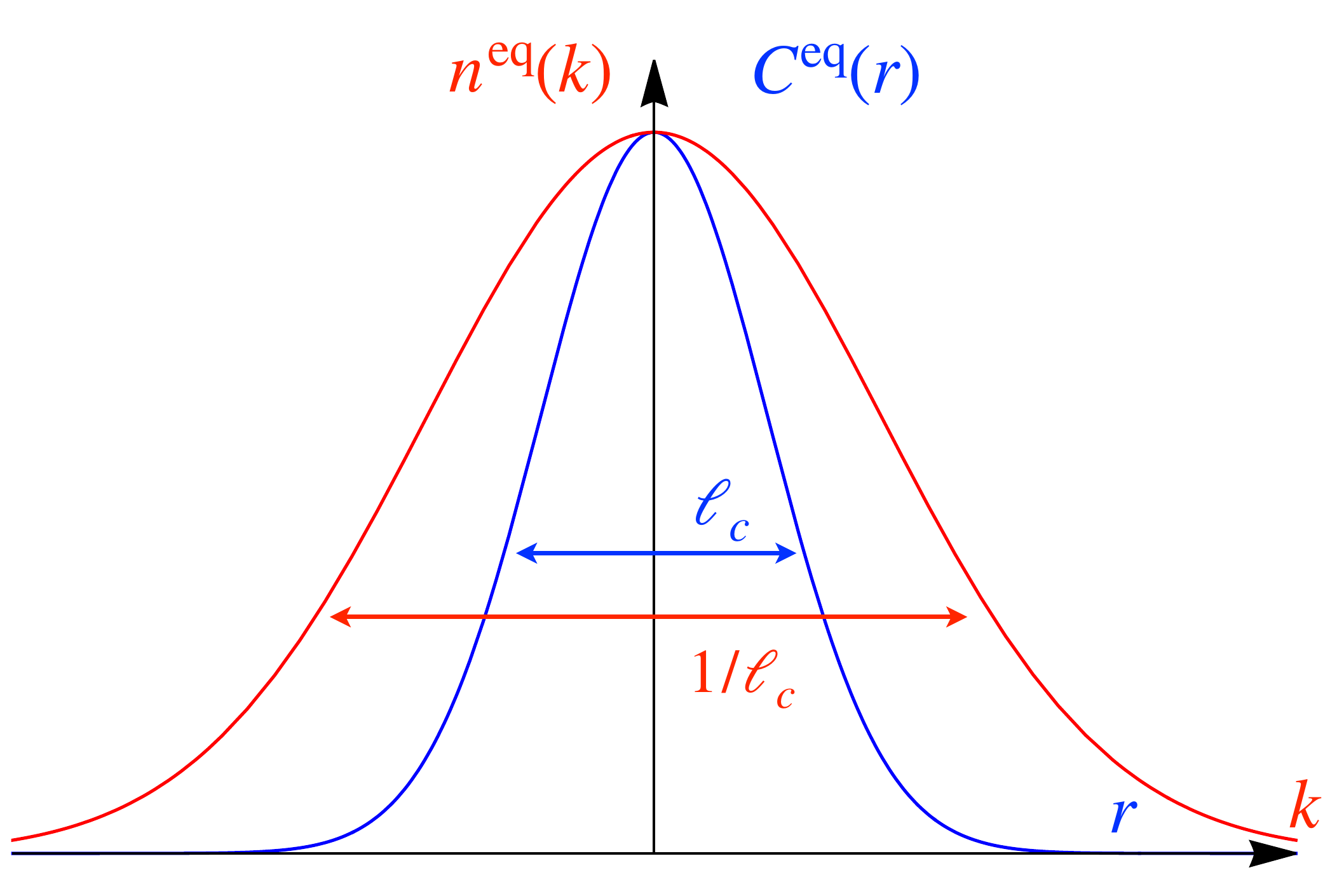}
\caption{Typical shape expected for the equilibrium two-point correlation function $C^{\rm eq}(r)$ we consider and the corresponding Wigner function $n^{\rm eq}(k)$.}
\label{fig:Typical-shape-expected}
\end{figure}

\subsection{Initial states and constraints on Wigner function}
\label{subsec:constraints} 

We are interested in the situation where half of the space is filled with particles. Therefore we focus on initial states of the form 
\begin{equation} \label{instate-continuous}
n(x,k,t=0)=n^{{\rm eq}}(k)\Theta(-x)
\end{equation}
where $\Theta$ is the Heaviside function and $n^{{\rm eq}}(k)$ is taken as the Wigner function of an equilibrium state (i.e., it is translationally invariant).
Being the Wigner function in one-to-one correspondence with the two-point correlation function, we can think of it as being associated to a class of states, rather than a single state (the correspondence becomes then unique if we also require the state to be Gaussian).
We will now put some restrictions on $n^{\rm eq} (k)$ in \eqref{instate-continuous} in order to illustrate our statement. 

First of all, to ensure that our Wigner function describes physical states, we consider $n^{{\rm eq}}(k)$ to be positive and bounded by one. This is in fact enough to ensure the same properties for the initial correlation function  $C^{{\rm eq}}(r) = \mathcal{F} [n^{\rm eq} (k)]$ (in turn needed to have occupations which are both positive and less than one). Note however that positivity is only a sufficient but non-necessary condition for the Wigner function, which in general can be also negative~\cite{wigner}. 

Moreover, we fix the local occupation number to the same value for every considered initial state leading to the condition $\int_\mathbb{R}\frac{\rmd k}{2\pi}n^{{\rm eq}}(k)=n_{0}$. In this way, we emphasize the role of coherence for transport, with no difference coming instead from the density imbalance. 

The actual restrictions are then the following. We assume $n^{{\rm eq}}(k)$ to be a ``bell-shaped'' function of $k$, namely symmetric with maximal value at $k=0$, and monotone (decreasing) on the two sides when going to higher values of $|k|$. Since $C^{{\rm eq}} (r)$ is given by the Fourier transform of $n^{\rm eq} (k)$, it will have the same qualitative bell-shape, and its typical correlation length scales inversely proportional to the width of $n^{{\rm eq}} (k)$. 
Finally, since we need all the conserved charges to be defined, and those are expressed as increasing powers of $k$, we further demand that the Wigner function must decay at least exponentially fast in $k$.

We show in Fig.~\ref{fig:Typical-shape-expected} the typical shape of the states we have in mind. Of course it does not encompass all
the possible states that one may consider (e.g., excited states corresponding in momentum space to multiple Fermi seas are not included by this picture). However, it includes states whose correlations decreases beyond a given coherence length, and states with smooth excitations above the Fermi surface. Those, in turn, describe a wide number of practical situations, such as e.g, finite temperature protocols, exponentially decaying two-point correlations, and so on (see Section~\ref{sec:quantitative}).

\subsection{The physical picture}

Under the assumptions outlined above, we want to argue that \emph{the more correlated the initial state, the slower the transport will be}.
We begin by giving qualitative arguments for this statement.

Since the density of the initial state is fixed on both sides for all protocols, it is not the density imbalance between $\mathcal{L}$ and $\mathcal{R}$ that generates differences in transport. This means that those are due to ``finer'' features of the Wigner function, while its integrated version is not enough: this is well expected since the model is integrable and standard hydrodynamics is not expected to hold in general. All the allowed initial states would lead to the same ``conventional hydrodynamic'' description, namely identical equations for density and hydrodynamic velocity, all starting from the same initial conditions. 

Because of the form of the initial state in Eq.~\eqref{instate-continuous}, the Wigner function of the $\mathcal{L}$-side,  $n^{\rm{eq}}(k)$, only depends on momentum, and its integral over $k$ is fixed to $n_0$. Therefore, the only freedom that we have is to redistribute the occupations of different $k$'s, while preserving the required bell-shape and without introducing gaps in momenta occupations.
Then, due to the Fourier correspondence between the Wigner function and the two-point function (associated to local correlations), it is natural that for correlations to extend over a large scale, the corresponding $n^{\rm{eq}}$ must be tight, and vice versa. But, for the class of states considered, this means that $n^{\rm{eq}}$ must be concentrated around small values of momenta. At the same time this means that particles in this configuration will spread at a slower pace. The link between correlations and transport properties is thus understood at a qualitative level.

\begin{figure}[t!]
\centering
\includegraphics[width=0.55\textwidth]{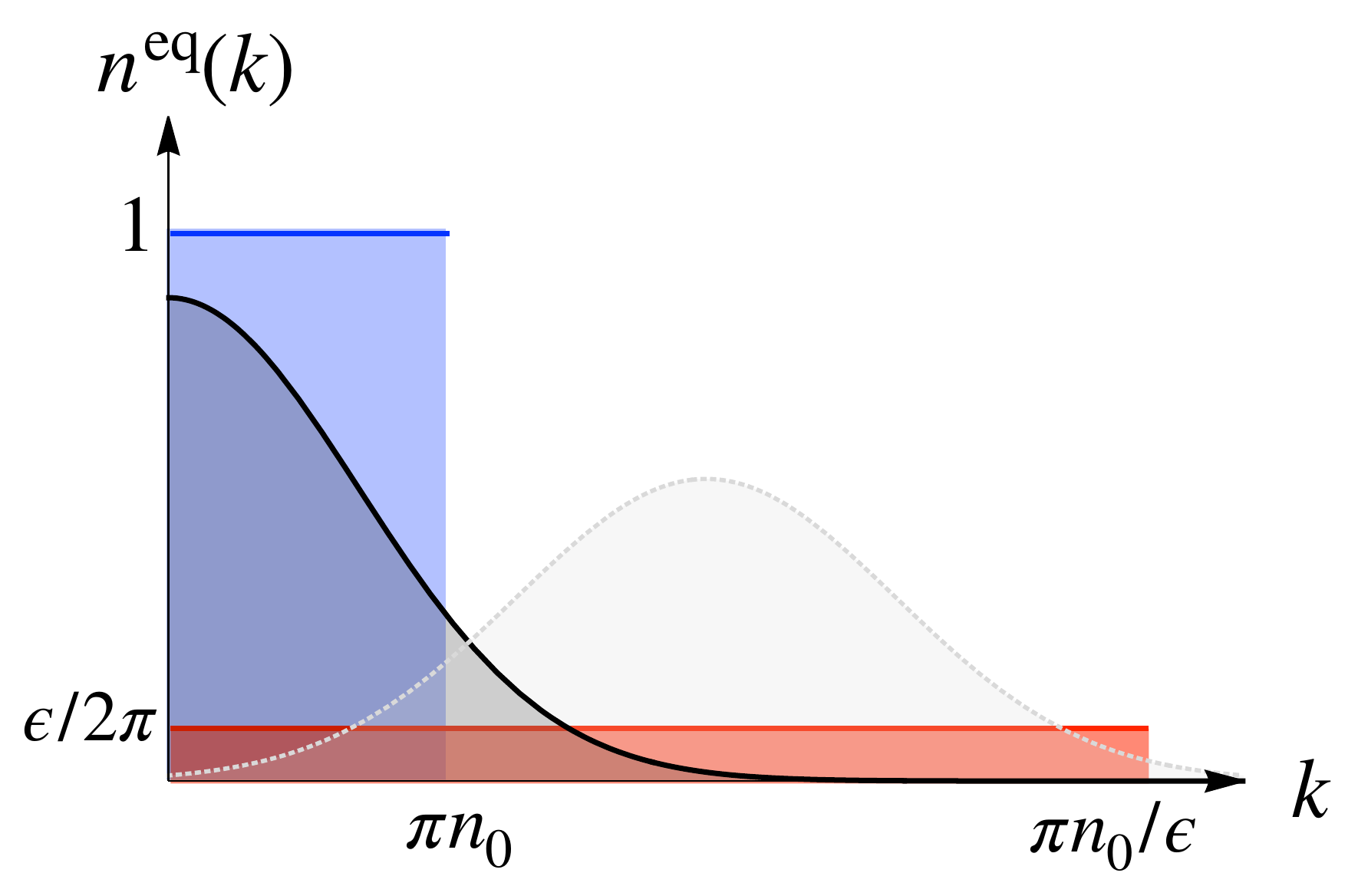}
\caption{Bell-shaped Wigner functions, which are the main focus of this work, are shown in blue, red and black (continuous lines). We only show $k>0$, making use of the parity of the considered class of $n^{\rm eq} (k)$. The two limiting cases given in Eqs.~\eqref{fermisea} and \eqref{fastest}, corresponding to long coherence length and slow transport in one case (blue), and short coherence length and fast transport in the other (red) are shown. Those correspond to the zero and infinite temperature case, respectively. A typical intermediate situation is shown in black. An example of a more general Wigner function, which does not fall in the same class, is also shown with a dotted line in light gray. All the Wigner function correspond to fixed total density $n_0$.}
\label{fig:qualitativepicture}
\end{figure}

This qualitative picture is actually enough to guess what the configurations corresponding to the extreme cases are. Those are illustrated in Fig.~\ref{fig:qualitativepicture}. In order to have slow transport, in fact, the optimum is the full occupation of low-energy modes: the associated Wigner function is such that 
\begin{equation} \label{fermisea}
n^{\rm{eq}}(k)=
\begin{cases}
1 & |k|<\pi n_0 \\
0 & \rm{otherwise}
\end{cases}
\end{equation}
However, this automatically gives the largest possible coherence length, being the tightest Wigner function we can construct with the given properties. Note that this is nothing but the single Fermi sea, namely the ground state of the system. Note also that this argument generalizes to generic fermionic integrable systems by replacing the Wigner function by the occupation number and is in agreement with inequality (2) of \cite{Doyon_lowerbound}.

On the other hand, to obtain fast particle transport, we aim to occupy high values of momenta. The optimum, under the required constraints, leads to 
\begin{equation} \label{fastest}
n^{\rm{eq}}(k)= \lim_{\epsilon \to 0} 
\begin{cases}
 \epsilon & |k|<\pi n_0/\epsilon \\
0 & \rm{otherwise}
\end{cases}
\end{equation}
which also coincides with the largest Wigner function we can have, therefore corresponding to the shortest range correlated state (the corresponding coherence length is actually zero) . This Wigner function is the one of an infinite temperature state. [Note that in the discrete case this is the equivalent of the protocol with no initial coherence, i.e., $\ell_c=1$.]

Releasing the constraints of Section~\ref{subsec:constraints}, one can find Wigner functions which are not in the class considered in this work. A simple example is to modify the Fermi sea in Eq.~\eqref{fermisea} by a shift to higher momenta, namely $n^{\rm{eq}}(k-k_0)$. This clearly does not change the correlations, since the corresponding two-point function just acquires a phase by Fourier transform. However, the corresponding state has faster particle transport, depending on the value of $k_0$. Similar reasoning applies to multiple Fermi seas and smoothed versions of those.
In such generalized situations, no unique link between correlations and transport properties can be established.
However, these states are not usually among the most natural choices as we will see with several examples, in particular at equilibrium. Nonetheless, they can easily appear in out-of-equilibrium settings: see for instance, inversion of population in pumped cavities \cite{opticalpumping} or quenches from the double(multi)-well potential~\cite{ruggiero2020quantum,ruggiero2021quantum}. 


\subsection{The quantitative picture} \label{sec:quantitative}
\subsubsection{Measuring transport ``speed'' and correlations}

So far, we have described our problem qualitatively. We now turn to a quantitative study. The questions we wish to address are the following: \textit{What do we mean exactly by ``fast'' or ``slow'' transport? How can we measure the amount of correlation in a given initial state?}
In the remaining of this Section, we will define two quantities with such precise meaning.

%

In order to quantify transport, let us focus on the number $N_{{\cal R}}(t)$ of particle in ${\cal R}$. As detailed in Appendix~\ref{App:transportmeasure}, this quantity scales linearly with time, with proportionality constant entirely determined by $n^{\rm eq} (k)$,
\begin{equation}
N_{{\cal R}}(t)=\mu_{T}(n^{{\rm eq}})t,
\end{equation}
with 
\begin{equation}
 \mu_{T}(n^{{\rm eq}}):=\int_{0}^{\infty}\frac{\rmd k}{2\pi}\, n^{{\rm eq}}(k)k.
\label{slopetransport}
\end{equation}
Such a linear relation is characteristic of ballistic transport. 
In our case, it holds \emph{at all times} (as follows by ignoring boundary effects around $x=0$).
Consequently, $\mu_T(n^{\rm eq})$ directly probes the total current from left to right in the quasi-stationary state within the light-cone. The slope $\mu_{T}$ has a simple interpretation: the contribution of a positive mode to the transferred
particles is simply equal to its velocity $k$ weighted by the mode
occupation number $n^{\rm eq}(k)$. 
Eq.~\eqref{slopetransport} defines the transport ``speed'' of the protocol. Moreover, from the limiting cases derived above, Eqs.~\eqref{fermisea} and \eqref{fastest}, one sees that $ \mu_T \in [{\pi n_0^2}/{4},\infty )$ which gives a lower bound on transport again consistent with Eq.~(2) of Ref.~\cite{Doyon_lowerbound}. 

Next, we need a quantity to characterize the amount of correlation
present in the equilibrium state $n^{{\rm eq}}(k)$. A natural quantity
to consider is 
\begin{equation} \label{C2}
\int_\R \big|C^{{\rm eq}}(r)\big|^{2}\rmd r.
\end{equation}
measuring how much any given point is correlated with all the rest of the system in the
initial state. This choice is further motivated by the fact that \eqref{C2} can be easily turned into a quantity
that depends on $n^{{\rm eq}}$. For translationally invariant states, in fact,
$n^{{\rm eq}}(k)$ is just the Fourier transform of $C^{{\rm eq}}(r)$, and
Parseval's equality then gives $\int_\R\big|C^{{\rm eq}}(r)\big|{}^{2}dr=\int\big|n^{{\rm eq}}(k)\big|{}^{2}\frac{\rmd k}{2\pi}$.
For convenience, however, instead of \eqref{C2} we consider the related quantity
\begin{equation} \label{muC}
\mu_{C}(n^{{\rm eq}}):=\int_\R n^{{\rm eq}}(k)(1-n^{{\rm eq}}(k))\frac{\rmd k}{2\pi}
\end{equation}
which is equal to $-\int_\R \big|C^{{\rm eq}}(r)\big|{}^{2}\rmd r$ up to a constant term, if we recall that $\int n^{{\rm eq}}(k)\frac{dk}{2\pi}= n_{0}$ for all initial states. Hence,
the higher $\mu_{C}$, the \emph{less correlated }is the state. We put $\mu_{C}$  under this precise form in order to compare it to other quantities such
as the purity (see Section~\ref{sec:ameasures}).

\subsubsection{Protocols}
\label{subsubsec:protocols}
\begin{figure}[t!]
\centering
\includegraphics[width=0.65\textwidth]{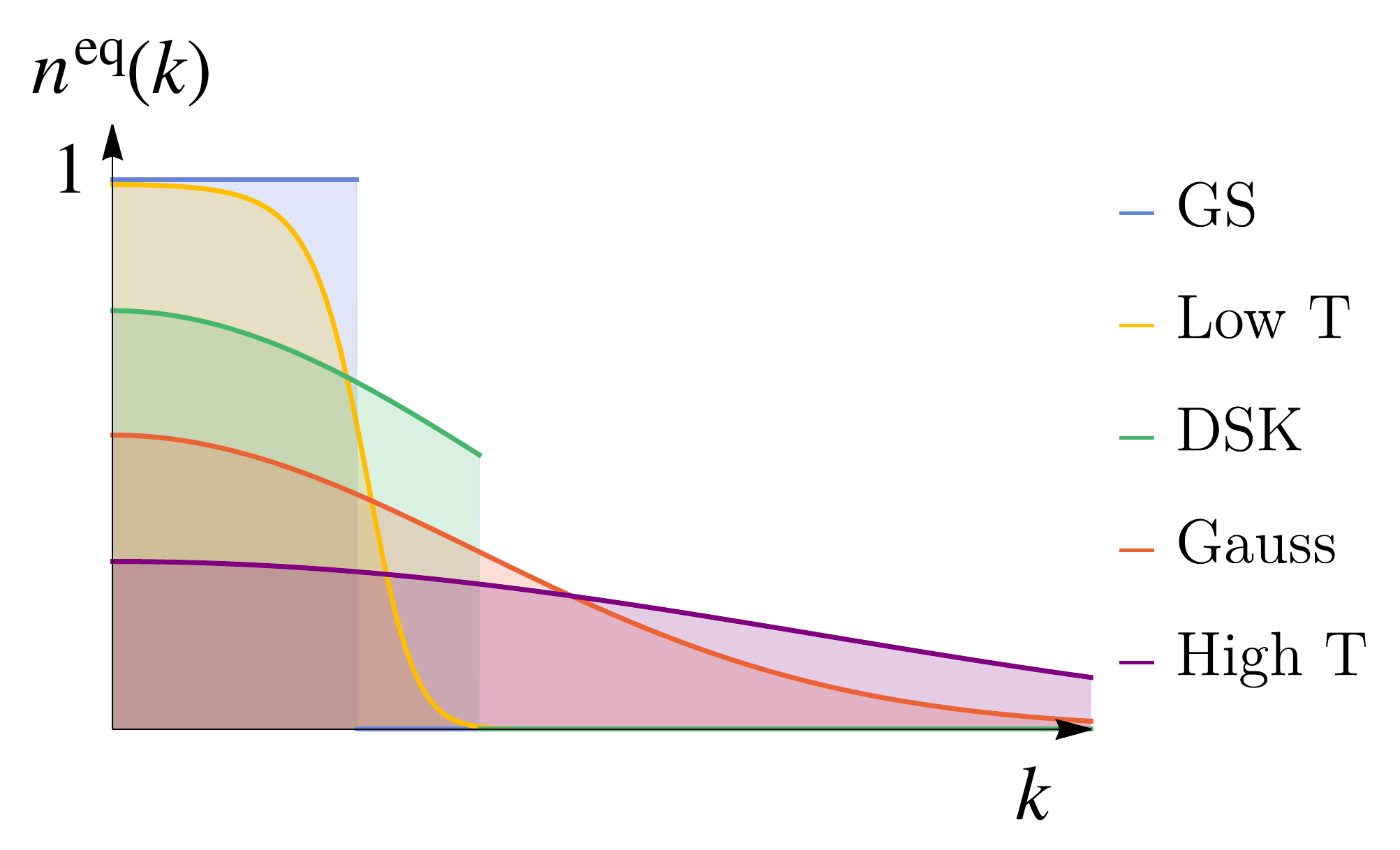}
\caption{Wigner functions corresponding to all the protocols considered in this work. Again we only plot $k>0$, using the parity of $n^{\rm eq}(k)$ for the class of ``bell-shaped'' states defined in Section~\ref{subsec:constraints}.}
\label{fig:Plots-of-the}
\end{figure}

We list below the different classes of initial states -- or protocols -- that we choose to study, whose properties are summarized in Table~\ref{table-protocols}. The associated Wigner functions are shown in Fig.~\ref{fig:Plots-of-the}.

\begin{itemize}

\item \textbf{Thermal}:
States in thermal equilibrium are a natural choice to start with, being the focus of most literature up to now~\cite{antal1998isotropic,Schutz_XXzerotemps,Eisler_FCS,ogata2002diffusion,gobert2005real,aschbacher2006out,PhysRevB.88.134301,karrasch2013nonequilibrium} (see also \cite{bertini2020finite} and references therein for recent literature). For these states, the occupation of modes is described by the
Fermi-Dirac distribution: 
\begin{equation} \label{finiteT}
n^{{\rm eq}}(k)=\frac{1}{1+e^{\beta(k^{2}/2-\mu)}}
\end{equation}
where the chemical potential $\mu$ is fixed by the implicit equation 
\begin{equation}
n_{0}=-\frac{1}{\sqrt{2\pi\beta}}{\rm Li}_{1/2}(-e^{\beta\mu})
\end{equation}
where ${\rm Li}_{s} (z)=\sum_{k=1}^{\infty}\frac{z^k}{k^s}$ is the polylogarithmic function,  such that
the normalization constraint is verified $\int_\R \frac{\rmd k}{2\pi}n^{{\rm eq}}(k)=n_{0}$. To our knowledge, there
is no analytical expression of the two-point correlation function,
and, for all practical computations, we use its definition as the inverse
Fourier transform of $n^{{\rm eq}}(k)$. 

The scaling of $C^{{\rm eq}}(r)$ with respect to $r$ is exponential for each $\beta <\infty$, while we recover the standard sine kernel behavior ($\sim 1/r$) in the ground state ($\beta=\infty$). However, at low enough temperature we expect it be very similar to the ground state behaviour: indeed, in the regime $1\ll r\ll \beta$, one observes that $C^{\rm eq}(r)$ behaves like $1/r$. 

\item \textbf{Gaussian}: Such states are defined from their two-point correlation function
\begin{equation} \label{gaussian}
C^{{\rm eq}}(r)=n_{0}e^{-\alpha r^{2}}
\end{equation}
where $\alpha$ is a free parameter that determines the typical distance on which correlations decay. For finite $\alpha$ this class contain states with \emph{finite} correlation length.

The corresponding Wigner function has gaussian form as well, and reads
\begin{equation}
n^{{\rm eq}}(r)=\sqrt{\frac{\pi}{\alpha}}n_{0}e^{-k^{2}/4\alpha}.
\end{equation}
\item \textbf{Deformed Sine-Kernel}: The deformed sine kernel (DSK) has been proposed in the literature to investigate the bulk of the spectrum of weak-orthogonal/Hermitian/symplectic random matrices~\cite{DeformedSine1,DeformedSine2,DeformedSine3}. The two-point correlation is given in this case by 
\begin{equation} \label{deformedSK}
C^{{\rm eq}}(r)=\frac{1}{{\cal N}}\int_{0}^{1}\rmd t\,\cos(\gamma r t)e^{-\sigma^{2}t^{2}}
\end{equation}
where $\gamma$ and $\sigma$ are free parameters and  ${\cal N}=\frac{\sqrt{\pi}}{2\sigma n_{0}}{\rm erf}(\sigma)$ the normalisation factor. Similarly to the ground state, it decays as a power law $\sim 1/r$. 
The corresponding Wigner function is given by a truncated Gaussian
\begin{equation}
n(x,k,t)=\begin{cases}
\frac{\pi}{{\cal N\gamma}}e^{-\big(\sigma\frac{k}{\gamma}\big)^{2}} & \text{for }|k|<\gamma\, ,\\
0 & \text{otherwise }.
\end{cases}
\end{equation}
\end{itemize}

\begin{table}
\centering
\begin{tabular}{|c|c|c|}
\hline 

\textbf{Protocol} & $C^{{\rm eq}}(r)$ & $n^{{\rm eq}}(k)$

\tabularnewline
\hline 
\hline 
Thermal & $\int\frac{dk}{2\pi}n^{{\rm eq}}(k)e^{ik(x-y)}$ & $\frac{1}{1+e^{\beta(k^{2}/2-\mu)}}$\tabularnewline
\hline 
Gaussian & $n_{0}e^{-\alpha r^{2}}$ & $\sqrt{\frac{\pi}{\alpha}}n_{0}e^{-k^{2}/4\alpha}$\tabularnewline
\hline 
Deformed sine kernel & $\frac{1}{{\cal N}}\int_{0}^{1}dt \,\cos(\gamma r t)e^{-\sigma^{2}t^{2}}$ & $\begin{cases}
\frac{\pi}{{\cal N\gamma}}e^{-\big(\sigma\frac{k}{\gamma}\big)^{2}} & \text{for }|k|<\gamma\\
0 & \text{otherwise }
\end{cases}$\tabularnewline
\hline 
\end{tabular} 
\caption{Classes of initial states (protocols) considered in this work.  \label{table-protocols}}
\par
\end{table}

\subsection{The transition map and its universality} \label{sec:transition_map}

\begin{figure}[t!]
\centering
\includegraphics[width=\textwidth]{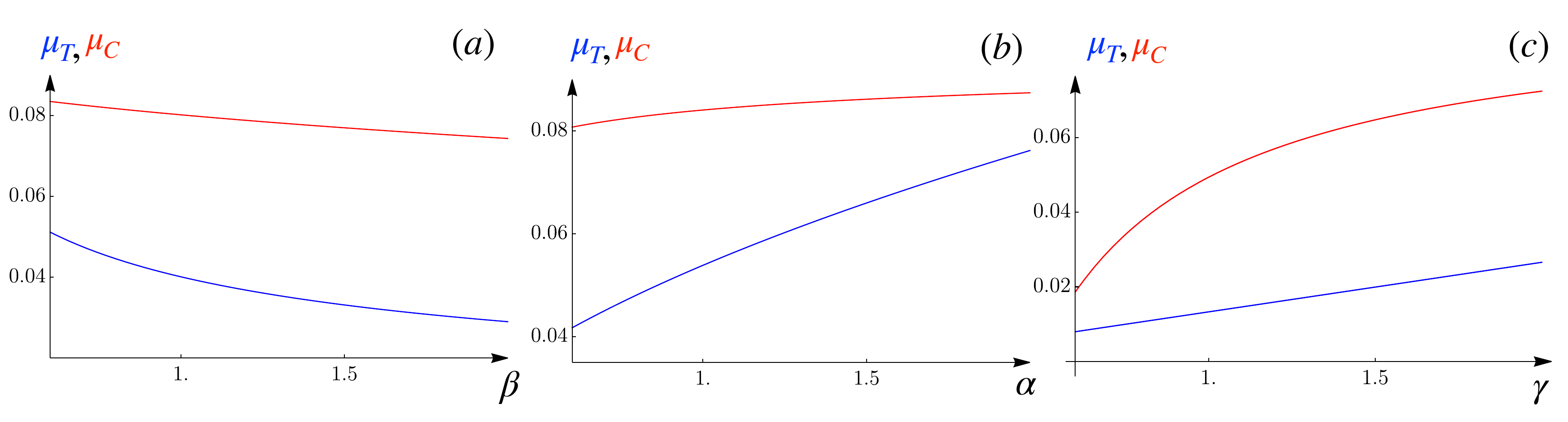}
\caption{Qualitative comparison of different protocols. Transport and correlations measures, $\mu_T$ and $\mu_C$ are shown as function of the different parameters. {\bf Panel (a):} Finite temperature protocol as a function of the inverse temperature $\beta$ (see Eq.~\eqref{finiteT}). {\bf Panel (b):} Gaussian protocol as a function of the parameter $\alpha$ (see Eq.~\eqref{gaussian}). {\bf Panel (c):} Deformed sine kernel protocol, with $\sigma=2$, as a function of the parameter $\gamma$ (see Eq.~\eqref{deformedSK}).}
\label{fig:Qualitative-comparison-of}
\end{figure}

Let us look at the dependence of our measure for transport $\mu_{T}$ in \eqref{slopetransport} on the
correlation measure $\mu_{C}$ in \eqref{muC}, for the three aforementioned protocols. 
The dependence of these measures on the model parameters are shown in Fig.~\ref{fig:Qualitative-comparison-of}.
We see that for all three one-parameter protocols $\mu_{T}$ and $\mu_{C}$ show the same monotonicity. This confirms our statement relating inital correlations with transport within each family of states.

However, we would like to address a harder question: is it possible to find a relation between $\mu_T$ and $\mu_C$ which is \emph{protocol-independent}?

\begin{figure}[t!]
\begin{centering}
\includegraphics[scale=0.6]{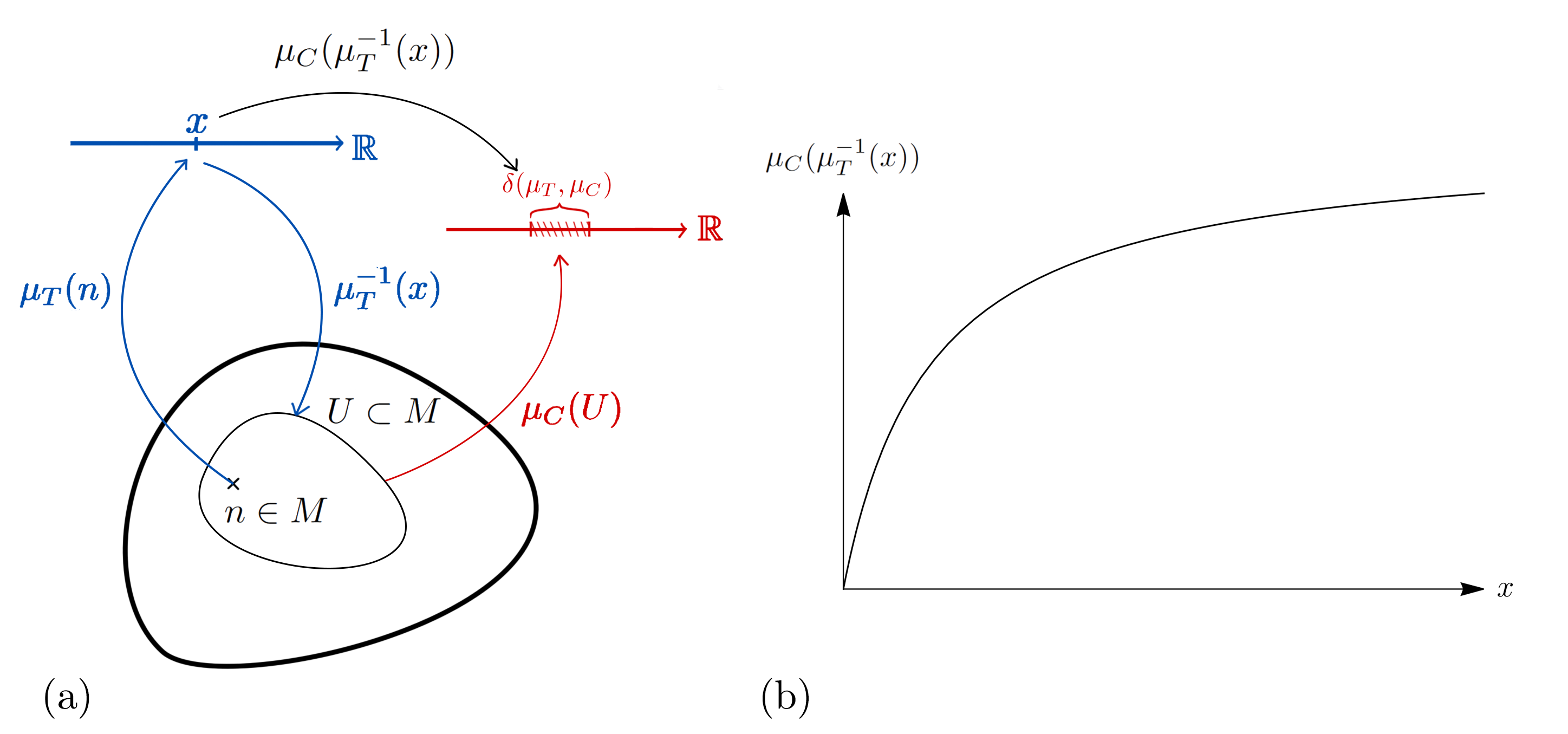}\caption{{\bf (a)} The transition map. $M$ is the set of acceptable Wigner functions, $n$ an element of it. $\mu_T$ and $\mu_C$  are applications from the set to $\mathbb{R}$. We define $U$ the preimage of $x\in\mathbb{R}$  by $\mu_T$. The map $\mu_{C}(\mu_{T}^{-1}(x))$ defines a transition map from $\mathbb{R}$ to $\mathbb{R}$ if the dispersion $\delta(\mu_T,\mu_C)$ of the set of points $\mu_C(U)$ is not too big. {\bf (b)} Plot of the transition map over the range of parameters for which $\delta \approx 10^{-4}$.}
\label{fig:The-transition-map}
\par\end{centering}
\end{figure}

To answer this question, we need to establish a more precise link between $\mu_T$ and $\mu_C$ from a mathematical point of view.
To this aim, let $M$ be the set of physically relevant Wigner functions as defined before.
Then $\mu_{T}$ and $\mu_{C}$ can be viewed as two applications from $M$ to
the set of real numbers $\mathbb{R}$. In principle, $M$ is an infinite
dimensional manifold, so that $\mu_{T}$ and $\mu_{C}$ are non-invertible.
Nevertheless, for any given function $\mu (x)$ and any $x\in\mathbb{R}$, one can define the subset $U$ of $M$ made of the preimages of $x$, i.e., $U:=\mu^{-1}(x)$. In general for any two functions $\mu_{1},\mu_{2} :M\to\mathbb{R}$,
there are no reasons for the set of points $\mu_{2}(\mu_{1}^{-1}(x))$ to
be related to one another. 
However if it turns out that, on the contrary,
the points $\mu_{2}(\mu_{1}^{-1}(x))$ are concentrated (up to a small dispersion) around
a given value, this defines a so-called \emph{transition map}. In our case, the reason why this dispersion should be small for the two functions $\mu_T,\mu_C$ lies in the qualitative discussion of the previous section, i.e., states with approximately the same transport properties have approximately the same correlations.  The transition map then provides
the relation between transport and correlations in a \emph{protocol-quasi-independent way}, where the ``quasi'' takes into account the small
dispersion of the set of points $\mu_{T}(\mu_{C}^{-1}(x))$.
This reasoning is summarized in Fig.~\ref{fig:The-transition-map}. 

We now proceed to evaluate the dispersion explicitly. Of
course, we do not have a complete parametrization of the physically
relevant Wigner functions. So, we restrict the
three protocols introduce above, i.e., finite temperature, Gaussian and DSK, which are explicitly parametrized by $\beta,\alpha,$ and ${\sigma,\gamma}$, respectively. We further set $\sigma=2$. We then assume that our set $M$ is given by
all the Wigner functions described by the three remaining parameters.
The procedure ensues as follows. For
$x\in\mathbb{R}$, the map $\mu_{T}^{-1}(x)$ gives us three elements
of $M$ corresponding to a tuple $(\beta,\alpha,\gamma)$. Then $\mu_{C}(\mu_{T}^{-1}(x))$
gives us three real points $\{y_{1},y_{2},y_{3}\}$. We
define the dispersion $\delta(\mu_T,\mu_C)$ between $\mu_{T}$ and $\mu_{C}$
as the averaged mean distance of these points to their centroid.  If
$\bar{y}={(y_{1}+y_{2}+y_{3})}/{3}$, and $\|y_{1}-y_{2}\|=\sqrt{(y_{1}-y_{2})^{2}}$, we have
\begin{equation} \label{deltaTC}
\delta(\mu_T,\mu_C):=\bigg\vert\frac{1}{N}\sum_{j=1}^{N}\frac{1}{3\bar{y}(x_{j})}(\|y_{1}(x_{j})-\bar{y}(x_{j})\|+\|y_{2}(x_{j})-\bar{y}(x_{j})\|+\|y_{3}(x_{j})-\bar{y}(x_{j})\|)\bigg\vert.
\end{equation}
where $x_{j}$ are our sample points in the image of $M$ by $\mu_{T}$
(we take it to be discrete in the example for computational reasons). This is nothing but the usual standard deviation. 
For the numerical application we choose $x_{j}\in[0.01,0.40]$ and
$x_{j+1}-x_{j}=0.01$ which corresponds to an inverse temperature
$10^{-2}  \lesssim \beta \lesssim 30$ in the finite temperature protocol. We then have that $\delta(\mu_T, \mu_C)\approx 10^{-4}$,  namely it is extremely small. 
As a consequence, the transition map $\mu_{C}(\mu_{T}^{-1}):\mathbb{R}\to\mathbb{R}$
is \emph{quasi-independent} of the intermediate point one
chooses on $M$ for the transition, where again, ``quasi'' means up to corrections of order $\delta(\mu_T,\mu_C)$. 
This is important as it implies that we may pick \emph{any} of our given protocols to compute it. For simplicity, we choose the Gaussian protocol, in which case we readily find the transition map as
\begin{equation}
\mu_{C}(\mu_{T}^{-1}(x))=n_{0}\left(1-\frac{n_{0}^{2}}{\sqrt{2}x}\right)
\end{equation}
where $x\in[\frac{n_{0}^{2}}{\sqrt{2}},\infty)$. The knowledge of this map then allows to compute transport properties from correlation
properties and vice versa. And since this is an increasing
monotonic function such that $\mu_T$ grows with $\mu_C$, we indeed verify that the \emph{ the less correlated the state is, the faster the transport will be}.

Let us now comment, however, on the fact that the transition map is not always well defined when the dispersion $\delta$ isn't small. For instance, notice that the previous function vanishes for $x=\frac{n_0^2}{\sqrt{2}}$. On the other hand, we know that for the finite temperature protocol this should occur at zero temperature, i.e., for $x=\frac{n_0^2\pi}{4}$. The two values differ approximately by $\approx 0.1 n_0^2$. This comes from the fact that $\delta(\mu_T,\mu_C)$ is not small anymore for very low temperatures ($\beta \gtrsim 10^2$) but rather of the order of $\approx 0.1$.  Note however that we do not see any discrepancy at high temperatures (we tested until $\beta \approx 10^{-4}$).

Importantly, the independence of the transition map with respect to
the protocol would be particularly useful in situations where one of $\mu_{T}$
or $\mu_{C}$ is difficult to compute while the other is easily accessible. One could then employ the transition map to deduce the unknown quantity from the known one.

\subsection{Alternative measures (and alternative statements)} \label{sec:ameasures}

We want to emphasize that the choice of the measures for transport and correlations is somewhat arbitrary. Indeed, other quantities could be used to measure transport as long as they are increased by the presence of large values of $k$, as $\mu_T$ does. Similarly, other measures of correlation are admissible if they favor values of $n$ close to $1$, as $\mu_C$ does.


For instance, we could replace $\mu_C$ by the \emph{purity} $\mu_P$, which for Gaussian states is defined as  
\begin{equation} \label{purity}
    \mu_{P}(n^{\rm eq}):=-\int\frac{\rmd k}{2\pi}\log(n^{\rm eq}(k)^{2}+(1-n^{\rm eq}(k))^{2}) \ .
\end{equation}
Again, the important point is then
to show that the dispersion $\delta(\mu_T, \mu_P)$ is small in order for the
transition map $\mu_{P}(\mu_{T}^{-1})$ to be defined in a protocol-independent way. For the same sampling points than before, we get again $\delta(\mu_T,\mu_P) \approx 10^{-4}$. For the transition map, we now get the formal expression: 
\begin{equation}
\mu_{P}(\mu_{T}^{-1}(x))=\int_\mathbb{R}\frac{\rmd k}{2\pi}\log \left( 1-\frac{2n_{0}^{2}}{x}e^{-\frac{n_{0}^{2}k^{2}}{4\pi x^{2}}}\left(1-\frac{n_{0}^{2}}{x}e^{-\frac{n_{0}^{2}k^{2}}{4\pi x^{2}}}\right)\right)
\end{equation}
which is still an increasing monotonic function of $x$. Thus, our
previous statement also translates for the purity for Gaussian density
matrices: \emph{the purer the state, the slower the transport is}.

\section{Generalization to the interacting case}\label{interacting}

In this Section we want to shortly comment on the complications arising when trying to export our statement to more general integrable models with interactions.

We start by noting that the dynamical equation for the Wigner function Eq.~\eqref{eq:continuityequation}, which is exact, coincides with the continuity equation satisfied by the {\it hydrodynamic} occupation function
\begin{equation}
    f_\mathrm{hydro}(x,k,t)=\frac{1}{1+e^{\beta^j(x,t)h_j(k)}}, \quad h_j(k)=\frac{k^j}{j!},
\end{equation}
where the summation over the repeated indices is understood \cite{GHD_Ben_Olalla_Takato}. The name of this function stems from the fact that the dependence on space and time is encoded through Lagrange multipliers $\beta^j(x,t)$, which is a consequence of the hydrodynamic approximation and is valid only on the hydrodynamic scale \cite{Spohn1991}. For free fermions in the continuum
the agreement of the dynamical equation implies that by demanding the matching of the initial conditions $\beta^j(x,0)=\frac{d^j}{dk^j}\log(f_\mathrm{hydro}(x,k,0)^{-1}-1)_{k\to0}$, we can identify the Wigner function with the occupation function at any time.

This observation motivates us to try to generalize our previous analysis to {\it interacting} integrable systems, where the notion of Wigner function is not necessarily obvious. First, the slope $\mu_T$ in the free case should now be replaced by the following:
\begin{equation}
    \mu_T=\int_0^\infty \rmd k\,v^\mathrm{eff}(k)\rho(k)=\frac{1}{2}\int_\mathbb{R} \rmd k\,|v^\mathrm{eff}(k)|\rho(k),
\end{equation}
where $\rho(k)$ and $v^\mathrm{eff}(k)$ are respectively the root density and the effective velocity of the system. The root density characterizes the distribution of quasi-particles of the system, while the effective velocity can be thought of as the average velocity of a quasi-particle when propagating on the hydrodynamic scale\footnote{Microscopically, the effective velocity is nothing but the sound velocity of the elementary excitations over a thermal or a GGE state \cite{PhysRevLett.113.187203}. This parallels the standard sound velocity over the Fermi sea.}. The precise definitions are not important here, and interested readers can refer, for instance, to \cite{GHD_Ben_Olalla_Takato,PhysRevLett.117.207201}. The interpretation of the slope also inherits from the non-interacting case.

Less straightforward is the correlation measure in the case of interacting integrable systems. Na\"ively one can replace $\mu_C$ with
\begin{equation}
    \mu_C[\rho]=\int_\mathbb{R} \rmd k \rho(k)(1-f_\mathrm{hydro}(k)),
\end{equation}
but now this quantity has no bearing on correlations anymore. Indeed, recall  first that  the occupation function $f_{\rm hydro} (k)$ cannot be in general expressed by correlation functions in a simple fashion as in the free case $n(k)=\langle \Psi^\dagger_k\Psi_k\rangle=\int_\mathbb{R} \rmd y e^{\I ky}\langle\Psi^\dagger(y/2)\Psi(-y/2)\rangle$. This is because while $k$ in the Wigner function is simply the momentum of the system, $k$ in the occupation function is the quasi-momentum of the system, namely the momentum of quasi-particle, corresponding to the modes which spread freely in the interacting case. To distinguish them, let us denote the latter by $\theta$ instead. Then the two objects, $k$ and $\theta$, are related by the Bethe equations, which is a set of many-body coupled equations in interacting integrable systems. Such a convoluted many-body relation between $k$ and $\theta$ in general makes it intractable to relate the momentum creation/annihilation operators $\Psi^\dagger_k$ and $\Psi_k$ with operators $Z^\dagger_\theta$ and $Z_\theta$ that create/annihilate a quasi-particle with quasi-momentum $\theta$ [note that they are known to satisfy the algebra called Zamolodchikov-Faddeev algebra \cite{Smirnov1992}]. Probably a more natural candidate for measuring correlation is the purity (i.e. the second R\'enyi entropy). However the replacement $n(k)/(2\pi)\mapsto\rho(k)$ is no longer valid, and it is known that in interacting integrable systems the purity is given in a more involved way \cite{Mesty_n_2018}.

The bottom line is that it seems a straightforward generalization of the relation between correlation and transport properties to general interacting integrable models is not possible. This is beyond the aim of our work, and is left to future investigation.
\section{Conclusions and perspectives}~\label{conclusion}
In this work we have established a direct link between equilibrium spatial correlations and transport properties of fermionic free systems. In particular, under the constraints mentioned in Section~\ref{subsec:constraints} we have shown that \textit{the more correlated a quantum state is, the slower its transport will be}. We first illustrated this statement in Section~\ref{sec:discreteCase} on an explicit example in the discrete setting, in which case the solution was obtained exactly. 
Building from the exact discrete illustration, we have provided a more general picture in the continuous limit in Section~\ref{sec:ContinuousCase}. A central object to this study was the Wigner function, which, at equilibrium is simply the Fourier transform of the two-point correlator for translation invariant states. We showed at the qualitative level that the relation between correlations and transport could be understood from the shape of the Wigner function for a subclass of physically relevant states. On the more quantitative side, we defined proper measures of correlations and transport ``speed'', and showed the existence of a \textit{transition map} which relates the correlation measure to the transport measure. Quite remarkably, this transition map is quasi-independent of the state for which it is considered over a large range of parameters.\\

Our study raises a certain number of question. The first is whether this link can be established for interacting systems which we briefly discussed in the last section. For integrable systems, generalized hydrodynamics provides the answer to the transport problem but in terms of the quasi-momenta of the system. The so-called inverse quantum scattering problem consisting in decomposing physical quantities such as two-point correlation functions in terms of these quasi-momenta is in general highly non-trivial and this problem deserves a study of its own \cite{Qinversescattering}.

Secondly, our study was restricted to ballistic transport. For some prototypical diffusive models such as the dephasing model \cite{dephasingznidaric, TJdephasingmodel, GHDdephasing} or the quantum symmetric simple exclusion process \cite{OpenQSSEP}, it is known that the main contribution to the transport is determined in mean by the local density, i.e the ``off-diagonal'' terms do not matter. It would be interesting to understand how the loss of coherence is linked with the change in transport properties as one goes further into the diffusive regime and/or how the initial coherence is still visible or not in the fluctuations around the mean state. In the same spirit, it would be interesting to try to make sense of our picture for localized states of matter \cite{RevMBL}.

Another interesting question concerns the local relaxation hypothesis within fluid cells usually taken as a necessary assumption in any hydrodynamic theory, and thus, in particular, in GHD. GHD indeed assumes the existence of fluid cells that after a given time (the local relaxation times) are well approximated by a generalized Gibbs ensemble (GGE). A priori, such local relaxation would lose information about non-local correlations. However in the free case, the GHD equation is the continuity equation for the Wigner function, and we saw that it gave the correct prediction for transport, no matter the range of correlations in the initial state. Thus, as was already noted in \cite{10.21468/SciPostPhys.8.3.048}, GHD gives the \textit{exact} correct prediction at the level of two-point correlation in the continuum for free systems without the need to assume relaxation within local fluid cells (note that we never needed to take into account the characteristic scale of such fluid cells). We plan to report on this point in more details in a subsequent work. 

Finally, it is tempting to think about the dispersion $\delta$ of the transition map as defining equivalence classes between different states. In some sense, it is a \textit{topological} property since it is stable under deformations of the Wigner function. It would be very interesting to push this reasoning further and try to characterize other classes of states within this approach. For instance one could look at how $\delta $ varies when there are ``holes'' in the Fermi sea (namely the case of multiple Fermi seas~\cite{fokkema2014split,eliens2016general,vlijm2016correlations}). As discussed before, we do not expect our qualitative reasoning relating transport and correlation to be valid. The expected higher values for the dispersion \eqref{deltaTC} which could be then interpreted as the occurrence of a crossover or even a ``phase transition''. At this level, however, this is very speculative, and needs further investigations. 

\paragraph{Acknowledgements} We thank Pierre Le Doussal and Benjamin Doyon for useful discussions and comments at the early stage of the project, as well as Vincenzo Alba for many useful comments and discussions. We also thank Benjamin Doyon and Jerome Dubail for feedbacks on the manuscript. TJ and PR acknowledge support from the Swiss National Science Foundation under division II. AK acknowledges support from ERC under Consolidator grant number 771536 (NEMO). TG acknowledges support from the ANR grant ANR-17-CE30-0027-01 RaMaTraF.

\bibliographystyle{unsrt}
\bibliography{bibliography}

\newpage
\appendix

\section{Calculations in the discrete protocol}
\label{App:DetailsDiscreteProtocol}

\subsection{Two-point correlation function of a general Dicke state}
\label{subApp0:DetailsDiscreteProtocol}

In this subsection we compute the two-point function of a generic fermionic Dicke state, of which Eq.~\eqref{Dicke-particular} is a particular case.
This state reads
\begin{equation}
\ket{\phi} =\frac{1}{\sqrt{\binom{D}{d}}}\sum_{\substack{{\cal C}\in \{ 0,1\}^D\\ \abs{\mathcal{C}}=d}}\ket{{\cal C}} 
\end{equation}
describing $d$ particles delocalized over $D$ contiguous sites of the lattice, with $ \mathcal{C}$ quantum state and $\abs{\mathcal{C}}$ its Hamming weight.
The sum hence denotes all the possible ways of arranging $d$
particles on $D$ sites symmetrically. The special case in Eq.~\eqref{Dicke-particular} of Section~\ref{subsec:DiscreteProtocol} corresponds to the choice
\begin{equation}
    D=\ell_c, \qquad d=n_0 \ell_c\, .
\end{equation}
Denoting by $\rho=\ket{\phi}\bra{\phi}$ the corresponding density matrix, we evaluate the two point function $\hat{c}^{(D,d)}$ with matrix elements $c^{(D,d)}_{jk} = \Tr(\rho c_{j}^{\dagger}c_{k})$. For this, we need to count the number of elements that give a non-zero contribution, and the problem is entirely combinatorial. 
\begin{itemize}
    \item For the diagonal terms $j=k$ we must have one-particle at site $j$, the remaining
being in whatever configuration. There are $\binom{D-1}{d-1}$
possibilities that verify this constraint, hence
\begin{equation}
 c^{(D,d)}_{jj}=  \frac{\binom{D-1}{d-1}}{\binom{D}{d}}=\frac{d}{D}
\end{equation}
The ratio $d/D$ is the average density as expected.
\item For the off-diagonal terms $j\neq k$, we need to fix two-elements:
the site $k$ must be occupied and the site $j$ must be empty. There are
$\binom{D-2}{d-1}$ configurations that verify this. Another complication arise in this case due to the signs arising from the fermionic combinatorics. In the following we obtain the correlations assuming for simplicity that $D=\ell_c\geqslant 3$ and  $d=n_{0}\ell_{c}\geq 2$. Let $j<k$, we want to evaluate 
\begin{equation}
c^{(D,d)}_{jk}=\frac{1}{\binom{D}{d}}\sum_{C}\Tr \left( c_{k}\ket{\cdots0_{j}\cdots1_{k}\cdots } \bra{ \cdots1_{j}\cdots 0_{k}\cdots } c_{j}^{\dagger}\right)
\end{equation}
where the index of the sum runs over all cross-product of states having an occupied site at position $k$ and an empty site at position $j$. There are $\binom{D}{d-1}$ of such cross-products but we have to distinguish between those which will give a positive or a negative sign. By the fermionic algebra rules, we have that 
\begin{align}
c_{k}\ket{\cdots0_{j}\cdots1_{k}\cdots}  & =(-1)^{\#\text{fermions between 1 and k}}\ket{\cdots0_{j}\cdots0_{k}\cdots} \\
c_{j}\ket{\cdots1_{j}\cdots0_{k}\cdots}  & =(-1)^{\#\text{fermions between 1 and j}}\ket{\cdots0_{j}\cdots0_{k}\cdots} 
\end{align}
The relevant quantities are therefore the number of occupied sites
between $j$ and $k$. Let $m$ this number, the overall sign is $(-1)^m$. The number of configurations having $m$ occupied sites between $j$ and $k$ is 
\begin{equation}
\abs{\hat{c}^{(D,d)}}_{\# =m}=\binom{D-(k-j+1)}{d-m-1}\binom{k-j-1}{m}
\end{equation}
Not all possible values of $m$ are possible in this parametrization, hence we next determine which values are allowed. We have to distinguish two cases: 

\begin{enumerate}
    \item If there is enough space outside of the interval $[j,k]$ to fill
all of the $d$ particles, $m$ can start at 0. Hence if $D-(k-j+1)\geq d-1$
\begin{equation}
 m\in [0,\min\{k-j-1,d-1\}]
\end{equation}
In this case, the two-point function reads
\begin{equation}
c^{(D,d)}_{jk}=\frac{1}{\binom{D}{d}}\sum_{m=0}^{\min\{k-j-1,d-1\}}(-1)^{m}\binom{D-(k-j+1)}{d-m-1}\binom{k-j-1}{m}
\end{equation}
\item If there is not enough space outside to fill all of the $d$ particles, i.e. $D-(k-j+1)<d-1$, then 
\begin{equation}
    m\in [d-1-(D-(k-j+1)),\min\{k-j-1,d-1\}]
\end{equation}
In this case, the two-point function reads
\begin{equation}
c^{(D,d)}_{jk}=\frac{1}{\binom{D}{d}}\sum_{m=d-1-(D-(k-j+1))}^{\min\{k-j-1,d-1\}}(-1)^{m}\binom{D-(k-j+1)}{d-m-1}\binom{k-j-1}{m}
\end{equation}
\end{enumerate}
\end{itemize}
The two-point function is of Toeplitz type, meaning that $c^{(D,d)}_{jk}=c^{(D,d)}_{j-k,0}$.

In the main text, we made the special choice $d=1,D=\ell_c$ and used the abbreviated notation  $c_{jk}^{\ell_c}:=c_{jk}^{(\ell_c,1)}$. In that case we have simply that $c_{j,k}^{\ell_{c}}={1}/{\ell_{c}}$ for all
$ (j,k)$ in $[1,\ell_{c}]\times[1,\ell_{c}]$.


\subsection{Time dependent density profile}
\label{subApp2:DetailsDiscreteProtocol}

In this Subsection, we prove the result \eqref{eq:solutionCdiscrete} the two-point function of the free fermionic system when the initial condition is a band matrix with coefficient
\begin{equation}
C_{j_1-j_2}:=(\hat{\mathcal{C}}_0^{\mathcal{L}})_{j_1,j_2}(t=0)
\end{equation}
Then, the correlation matrix at time $t$ is :
\begin{eqnarray}
&&C_{m,n}(t) = (J N_0 J^\dagger)_{m,n}  \label{eq:beg}\\
&& \quad = \I^{m-n} \sum_{j_1,j_2 \leq 0} \I^{j_2-j_1} J_{m-j_1}(t) (\hat{\mathcal{C}}_0^{\mathcal{L}})_{j_1,j_2}  J_{n-j_2}(t) \\
&& \quad = \I^{m-n} \sum_{j_1,j_2 \leq 0} \I^{j_2-j_1} J_{m-j_1}(t) C_{j_1-j_2}  J_{n-j_2}(t) \\
&& \quad =  \I^{m-n} \sum_{k \in \mathbb{Z}}\I^k C_k  \sum_{j_1\leqslant \min(k,0)}   J_{m-j_1}(t)  J_{n-j_1+k}(t) \\
&& \quad =\I^{m-n} \sum_{k \in \mathbb{Z}} \I^k C_k B_t(m-(k)_-, n+k -(k)_-)\\
&& \quad =\I^{m-n} \sum_{k \in \mathbb{Z}} \I^k C_k B_t(m-(k)_-, n +(k)_+)
 \label{eq:end}
\end{eqnarray}
where we have introduced $(k)_- = \min(k,0)$, $(k)_+=\max(0,k)$ to ensure both the $j_1\leq 0$ and the $j_2 = j_1 -k  \leq 0$ conditions at once; and
\begin{equation}
\label{eq:BesselDisKer}
B_t(m,n)=\frac{t}{2(m-n)}\left(J_{m-1}(t)J_n(t)-J_m(t)J_{n-1}(t) \right).
\end{equation}
is the discrete Bessel kernel.
In particular, for the particle density (i.e., $m=n$), we obtain
\begin{equation}
\begin{split}
    C_{m,m}(t)&= \sum_{k \in \mathbb{Z}} \I^k C_k B_t(m-(k)_-, m +(k)_+)\\
    &= n_0 B_t(m,m)+\sum_{k> 0}\I^k (C_k+(-1)^k \bar{C}_k)B_t(m+k,m)\\
    & =n_{0}B_{t}(m,m)+2\sum_{k=1}^{\ell_{c}}\cos(\frac{k\pi}{2})C_{k}B_{t}(m+k,m)
    \end{split}
\end{equation}
where we made use of the fact that the correlation matrix is real in our case of interest.

\subsection{Hydrodynamic limit of density profile and integrated current}
\label{subApp3:DetailsDiscreteProtocol}

In this Subsection, we will prove the formula \eqref{eq:krajDensityLargeTime} and  \eqref{eq : discreteslope} for the density profile and the integrated current in the large time limit $t\gg \ell_c$. 
More precisely, we are interested in the hydrodynamic limit, namely the limit of $m,t \to \infty$, with $u =m/t$ fixed that we named $\lim_{\rm{h}}$. 

Let $\Phi$ denote the local density $C_{m,m} (t)$ in such hydrodynamic limit.
Its derivative with respect to $u$ is given by
\begin{equation}
\begin{split}
    \Phi '\left(u= \frac{m}{t} \right)& = \lim_{\rm{h}} \left(t [C_{m+1,m+1}(t)-C_{m,m}(t)]\right)\\
    &=\lim_{\rm{h}} \left(-t J_m(t)[n_0 J_m(t)+2\sum_{k=1}^{\ell_c}\cos(\frac{k \pi}{2})C_k J_{m+k}(t)]\right)\\
    \end{split}
    \label{eq:densityHydroDiff}
\end{equation}

\begin{figure}[t!]
    \centering
    \includegraphics[scale=0.8]{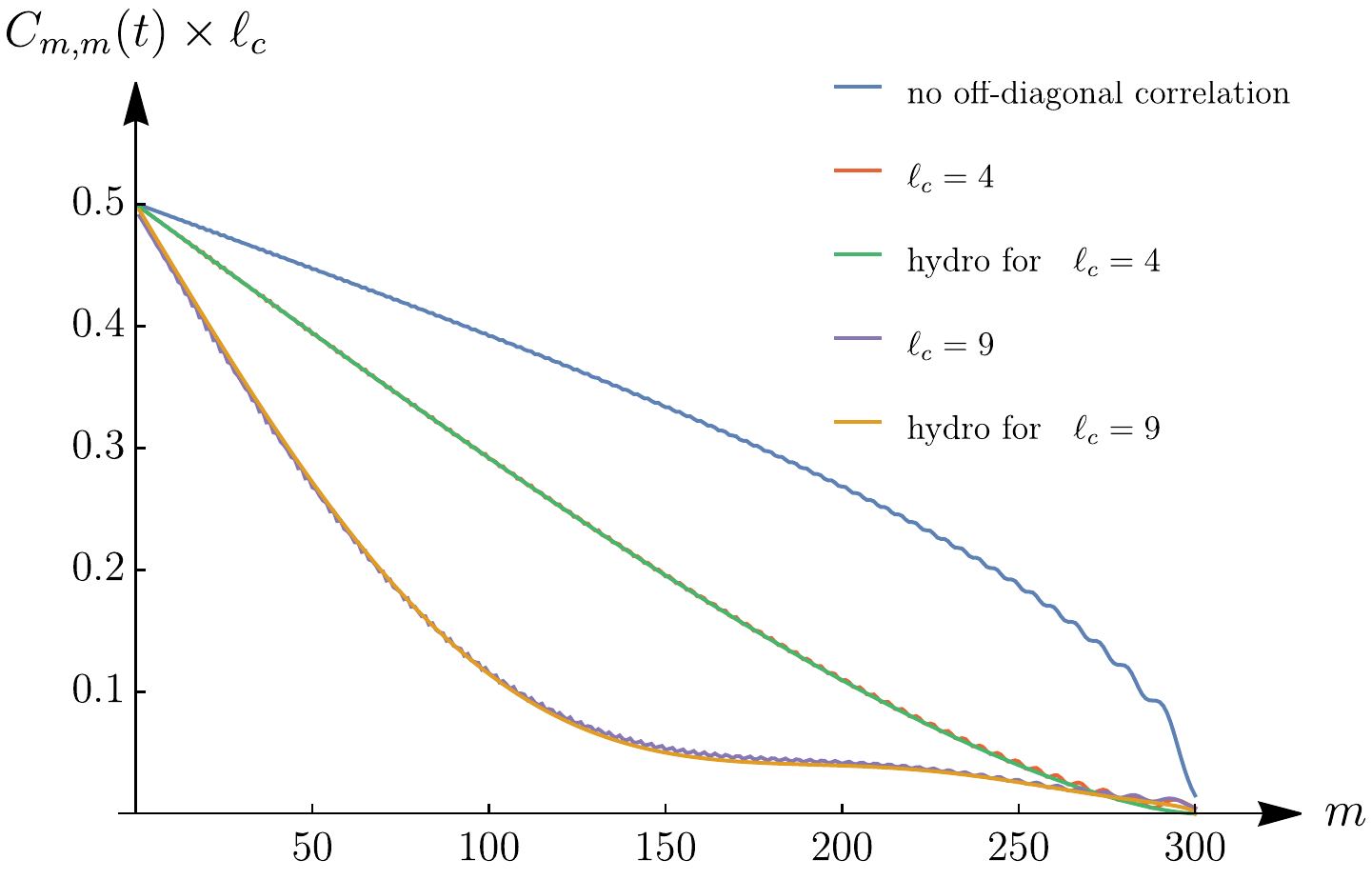}
\caption{Density profile at $t=300$ in the right half-space for difference
coherence length. For readability, all the curves have been rescaled
by a factor $\ell_{c}$ to match at the origin.}
\label{fig:discrete_coherenceApp}
\end{figure}

Now we take the limit of the Bessel function when its argument scales as its index
\begin{equation}
\label{asymptotics22}
J_{ut} (t) \sim \sqrt{\frac{2}{ \pi t \sqrt{1-u^2}  }}\cos(t ( \sqrt{1-u^2}-u\arccos u)-\frac{\pi}{4})
\end{equation}
and consider the Dicke state where $n_0=1/\ell_c$. Replacing this asymptotics in \eqref{eq:densityHydroDiff} and averaging oscillating terms, we obtain
\begin{equation}
\begin{split}
  &  \Phi'(u=\frac{m}{t})  =\lim_{\rm{h}} \left(-\frac{ 1}{\ell_c\pi}\frac{1}{\sqrt{1-u^2}}-
    \frac{2t}{\ell_c} J_m(t)\sum_{k=1}^{\ell_c}\cos(\frac{k\pi}{2})\frac{\ell_c-k}{\ell_c} J_{m+k}(t)\right)\\
    &=\lim_{\rm{h}}  \bigg(-\frac{ 1}{\pi \ell_c}\frac{1}{\sqrt{1-u^2}}-
    \frac{2t}{\pi} \frac{1}{(1-u^2)^{1/4}} \sum_{k=1}^{\ell_c}\frac{\ell_c-k}{\ell_c^2}\frac{\cos(\frac{k\pi}{2})}{(1-(u+\frac{k}{t})^2)^{1/4}} \\
    & \times \cos\left(t ( \sqrt{1-u^2}-u\arccos u- \sqrt{1-(u+\frac{k}{t})^2}+(u+\frac{k}{t})\arccos (u+\frac{k}{t}))\right)\bigg)\\
    \end{split}
\end{equation}
The only terms contributing are the ones with $k/t\ll 1$ (the others are irrelevant to the oscillations, which dominate). Hence, taking $t\gg \ell_c$ as our hydrodynamic limit, we have
\begin{equation}
\begin{split}
  &  \Phi'(u) =-  \frac{ 1}{\ell_c \pi}\frac{1}{\sqrt{1-u^2}}-
    \frac{2}{\pi }  \sum_{k=1}^{\ell_c}\cos(\frac{k\pi}{2}) \frac{\ell_c-k}{\ell_c^2} \frac{\cos\left(k\arccos(u)\right)}{\sqrt{1-u^2}}\\
    \end{split}
\end{equation}
Upon integration, the density reads
\begin{equation}
\begin{split}
    \Phi(u)&=\int_0^u \mathrm{d}v \, \Phi'(v)= \frac{\arccos(u)}{\pi \ell_c}+
    \frac{2}{\pi }  \sum_{k=1}^{\ell_c}\cos(\frac{k\pi}{2}) \frac{\ell_c-k}{k \ell_c^2} \sin(k \arccos(u))
    \end{split}
    \label{eq:AppDensityCotinuous}
\end{equation}
Integrating the density leads to the determination of the total number of particles on the right of our lattice in the hydrodynamics description
\begin{equation}
\begin{split}
    \mathcal{N}_{\cal R}(t) =t \int_0^1 \mathrm{d}u \, \Phi(u) = t \left(\frac{1}{\pi \ell_c }- \frac{2}{\pi \ell_c^2}  \sum_{k=2}^{\ell_c}\cos(\frac{k\pi}{2})^2 \frac{\ell_c-k}{k^2-1}\right)
    \end{split}
\end{equation}
The second term in the parenthesis is clearly negative, providing that the presence of a correlation length slows down the transport. This effect is displayed in Fig.~\ref{fig:discrete_coherenceApp} where we compute the particle density profile at $t=300$ and compare the discrete sum \eqref{eq:solutionCdiscrete} to the continuous description \eqref{eq:AppDensityCotinuous} which match in great accuracy.

\section{Full counting statistics in domain-wall states of lattice free fermions}
\label{subsec:FCS}

In this appendix, we give an expression of the full counting statistics in the discrete domain wall in terms of a continuous kernel for a general correlated domain wall initial state, thus extending the recent results in [Moriya, Nagao and Sasamoto, JSTAT 2019(6):063105] on the one-dimensional XX spin chain.

The (homogeneous) FCS is defined for a subspace $\mathcal{S}$ as the generating function of the particle number operator in $\mathcal{S}$ at time $t$ i.e. $\hat N_{\mathcal{S}}(t)=\sum_{m\in \mathcal{S}}c_m^\dagger(t)c_m(t)$. It is by Wick's theorem equal to a Fredholm determinant \cite{Nazarov,Schonhammer,Eisler_FCS,Sasamoto_LargedeviationXX}
\begin{equation}
\big\langle e^{\I \lambda \hat N_{\mathcal{S}}(t) }\big\rangle=\mathrm{Det}\left[ 1+(e^{\I \lambda}-1)\hat{C}(t)P_{\mathcal{S}}\right]_{\ell^2(\mathbb{Z})}
\end{equation}
where $P_{\mathcal{S}}$ is the projector onto the subspace $\mathcal{S}$, and $\ell^2 (\mathbb{Z})$ denotes the space of square-summable sequences on $\mathbb{Z}$. The FCS has been used in \cite{Eisler_FCS,Sasamoto_LargedeviationXX} to make a very interesting connection between the statistics of free fermions and the statistics of eigenvalues in random matrix theory. In a larger scope, we note that connections with random matrix theory and stochastic systems also exist for other fermionic systems, namely non-interacting fermions in a range of classic potentials, see for instance Ref.~\cite{dean2019noninteracting,NoncrossingBrownians,MatrixKestenFermionsMorse}.

In the following we will denote $z=e^{\I\lambda}-1$ and drop the time dependence in the kernels. We inject the expression of $\hat C$ given in Eq.~\eqref{eq:solutiondiscrete} in terms of the initial correlations and the discrete kernel $\hat{J}$.
We then use the general relation $\mathrm{det}(I+AB)=\mathrm{det}(I+BA)$ for arbitrary Hilbert-Schmidt operators $A$ and $B$ (this is the Sylvester's identity, valid as long as the different products make sense \cite{OperatorTheoryBook}). This yields
\begin{equation}
\mathrm{Det}\left[ 1+z\hat{J}^\dagger \hat{C}_0 \hat{J} P_{\mathcal{S}}\right]_{\ell^2(\Z)}=\mathrm{Det}\left[ 1+z\hat{C}_0 \hat{J} P_{\mathcal{S}}\hat{J}^\dagger \right]_{\ell^2(\Z)}
\end{equation}
Choosing the subspace $\mathcal{S}$ to be an interval $[a,b]$ with $0<a < b$, the operator $\hat{J} P_{\mathcal{S}}\hat{J}^\dagger$ is shown in Appendix~\ref{app:ChristoffelDarboux} to become
\begin{equation} 
\label{eq:ChristoffelDarbouxResult}
[\hat{J} P_{\mathcal{S}}\hat{J}^\dagger]_{n,m} =\I^{n-m}(B_t(m-b,n-b)-B_t(m-a+1,n-a+1)) 
\end{equation}
where $B_t$ is the discrete integral operator acting on $\ell^2(\Z)$ defined by the discrete Bessel kernel introduced previously (cf. Eq.~\eqref{eq:BesselDisKer}).

If we now assume that the interval is semi-infinite, i.e. $b=+\infty$, we have
\begin{equation}
\label{eq:BesselDisKerInf}
[\hat{J} P_{\mathcal{S}}\hat{J}^\dagger]_{n,m}=- \I^{n-m}B_t(m-a+1,n-a+1) := - \mathcal{B}_{n,m}
\end{equation}
where we have introduced the notation $\mathcal{B}$ for this discrete kernel, which depends on $t$ and $a$.
Hence we obtain the full counting statistics on an arbitrary interval for a general initial condition in terms of the discrete Bessel kernel as
 \begin{equation}\label{eq:general_FCS}
\big\langle e^{\I\lambda \hat N_{\mathcal{S}}(t) }\big\rangle=\mathrm{Det}\left[ 1- z \, \hat{C}_0 \, \mathcal{B}\right]_{\ell^2(\Z)}
\end{equation}
The form of this result Eq.~\eqref{eq:general_FCS} and in particular the operator $\mathcal{B}$, does not depend on the details of the initial condition, appearing only as the initial correlation matrix $\hat{C}_0$. \\

The integral operators we obtained so far acted on the lattice $\ell^2(\Z)$. Moriya, Nagao and Sasamoto recently obtained in \cite{Sasamoto_LargedeviationXX} that for the domain wall initial condition, the FCS $\big\langle e^{\I\lambda \hat N_{\mathcal{S}}(t) }\big\rangle$ can be seen as the determinant of the continuous Bessel kernel, acting on $\mathbb{L}^2(\mathbb{R^+})$, see Ref.~\cite[Eq.~(3.8)]{Sasamoto_LargedeviationXX}. This relation allowed them to make a connection with the Laguerre ensemble in random matrix theory and study the large deviations of the spin current. Here, we extend this result for \emph{any} initial condition $\hat{C}_0$ supported on the left-side $\mathcal{L}$ and zero elsewhere,
while we recall that the full counting statistics are studied on the subspace $\mathcal{S}=[a,+\infty)$ with $a>0$. In this view, we introduce the following semi-discrete kernels
\begin{equation}
\begin{split}
&\tilde{J}_1(n,\tilde{t})=\left(\frac{\I t }{\sqrt{\tilde{t}}}\right)^{n-a}  J_{n-a+1}(\sqrt{\tilde{t}}) \; ,\\
&\tilde{J}_2(\tilde{t},m)= \left(\frac{\I t}{\sqrt{\tilde{t}}}\right)^{a-m}  J_{m-a}(\sqrt{\tilde{t}})\; .
\end{split}
\end{equation}
where $m$ and $n$ denote integers and $\tilde{t}$ denotes a real variable (the differing order of variables is chosen in order to write operator products in a natural way). Crucially, it is shown in Appendix~\ref{app:DiscreteToContinuous} that the following discrete-operator identity holds for every matrix element $(n,m)$
\begin{equation}
\label{eq:KernelRelation}
    \hat{J}^\dagger P_{\mathcal{S}}\hat{J}=-\tilde{J}_1 \tilde{J}_2
\end{equation}
where the measure of integration in $\tilde t$ space is $\frac{1}{2} \int_0^{t^2} \frac{\rmd \tilde{t}}{\sqrt{\tilde{t}}}$, on the interval $[0,t^2]$. A straightforward application of Sylvester's identity then gives
\be 
\mathrm{Det}\left[ 1 - z \hat{C}_0 \tilde{J}_1 \tilde{J}_2\right]_{\ell^2(\Z)} =\mathrm{Det}\left[ 1 - z\tilde{J}_2 \hat{C}_0\tilde{J}_1 \right]_{\mathbb{L}^2(0,t^2)}
\ee
such that finally
\begin{equation} 
\label{eq:ContinuousToDiscrete}
\big\langle e^{\I \lambda \hat N_{\mathcal{S}}(t) }\big\rangle
=
\mathrm{Det}\left[ 1+z\hat{J}^\dagger \hat{C}_0 \hat{J} P_{\mathcal{S}}\right]_{\ell^2(\Z)}=\mathrm{Det}\left[ 1 - z\tilde{J}_2 \hat{C}_0\tilde{J}_1 \right]_{\mathbb{L}^2(0,t^2)} \; .
\end{equation}

If the initial state is the uncorrelated domain wall, $\hat{C}_0=\hat{C}_0^\mathcal{D}$ is the identity matrix when restricted to $\mathcal{L}$, and zero otherwise. In this case,
$J^\dagger \hat{C}_0^\mathcal{D} J P_\mathcal{S}$ is easily shown to be the discrete Bessel kernel through computations similar to the ones that lead to Eq.~\eqref{eq:ChristoffelDarbouxResult}. Furthermore, $-\tilde{J}_2 \hat{C}_0^\mathcal{D}\tilde{J}_1$ is the continuous Bessel kernel as shown in Appendix~\ref{app:RecoverContinuousBessel}. 
As a consequence, Eq.~\eqref{eq:ContinuousToDiscrete} recovers the Moriya et al. identity connecting the discrete Bessel kernel to the continuous Bessel kernel. We note that our proof of this identity is substantially shorter than the one given in \cite{Sasamoto_LargedeviationXX}, which is centered on the computation of all moments of both kernels in order to establish the identity at the level of Fredholm determinants. In contrast, the introduction of semi-discrete kernels $\tilde{J}_{1,2}$ and the use of Sylvester's identity allows us to avoid computations of arbitrary-order moments.

With a general initial condition, Eq.~\eqref{eq:ContinuousToDiscrete} is a generalization of the Moriya et al identity, Ref.~\cite[Eq.~(3.8)]{Sasamoto_LargedeviationXX}, connecting a discrete kernel to a continuous kernel that depends on the initial condition. An interesting development of this study would be to find XX chain initial states leading, through Eq.~\eqref{eq:ContinuousToDiscrete}, to other known random matrix theory kernels.


\subsection{Derivation of Eq.~\eqref{eq:ChristoffelDarbouxResult}}
\label{app:ChristoffelDarboux}
 
In this subsection, we establish the Christoffel-Darboux identity for the discrete Bessel Kernel given in Eq.~\eqref{eq:ChristoffelDarbouxResult}. We first write the following relation for the Bessel functions
\begin{equation}
\begin{split}
J_{n-j}(t)J_{m-j}(t)&=\frac{t}{2(m-n)}\left(\frac{2}{t}[m-j-n+j]J_{n-j}(t)J_{m-j}(t)\right)\\
&=\frac{t}{2(m-n)}\left(\frac{2}{t}[m-j]J_{n-j}(t)J_{m-j}(t)-\frac{2}{t}[n-j]J_{n-j}(t)J_{m-j}(t)\right) \; .
\end{split} 
\end{equation}
The recurrence relation of Bessel functions $(2 \nu / z) J_{\nu}(z)=J_{\nu+1}(z)+J_{\nu-1}(z)$ \cite{NIST:DLMF} yields
\begin{equation}
\begin{split}
J_{n-j}(t)J_{m-j}(t)&=\frac{t}{2(m-n)}\left(J_{n-j}(t)(J_{m-j+1}(t)+J_{m-j-1}(t))-(J_{n-j+1}(t)+J_{n-j-1}(t))J_{m-j}(t)\right) \\
&=\frac{t}{2(m-n)}\big(\left[J_{n-j}(t)J_{m-j+1}(t)-J_{n-j-1}(t)J_{m-j}(t)\right] \\ &   \qquad \qquad \qquad +\left[J_{n-j}(t)J_{m-j-1}(t)-J_{n-j+1}(t)J_{m-j}(t)\right]\big) \; .
\end{split}
\end{equation}
This is the general term of a telescopic sum leading to the simplified equation 
\begin{equation}
\begin{split}
\sum\limits_{j=a}^{b}J_{n-j}(t)J_{m-j}(t)=\frac{t}{2(m-n)}&\big(\left[J_{n-a}(t)J_{m-a+1}(t)-J_{n-b-1}(t)J_{m-b}(t)\right] \\ &+\left[J_{n-b}(t)J_{m-b-1}(t)-J_{n-a+1}(t)J_{m-a}(t)\right]\big) \; .
\end{split}
\end{equation}
With the discrete Bessel kernel $B_t(m,n)$ defined in the main text, we finally recover Eq.~\eqref{eq:ChristoffelDarbouxResult} of the main text
\begin{equation}
\sum\limits_{j=a}^{b}J_{n-j}(t)J_{m-j}(t)=B_t(m-b,n-b)-B_t(m-a+1,n-a+1) \; .
\end{equation}

\subsection{Derivation of Eq.~\eqref{eq:KernelRelation}}
\label{app:DiscreteToContinuous}

In this subsection, we establish the relation between discrete and semi-discrete kernels stated in Eq.~\eqref{eq:KernelRelation}. We will directly write the matrix element $(n,m)$ of $J^\dagger P_{\mathcal{S}}J$ and transform it using the contour integral expression of $J$ as $J_k(t) =\oint_{|z|=1}  \frac{\rmd z}{2\I \pi z} e^{\frac{ t}{2}(z-\frac{1}{z})} (-z)^k $, see  Ref.~\cite[equations A.6-A.8]{Sasamoto_LargedeviationXX}. We have:
\begin{equation}
\begin{split}
&[\hat{J}^\dagger P_{\mathcal{S}}\hat{J}]_{n,m}=\I^{n-m} \sum_{k=a}^b J_{n-k}(t)J_{m-k}(t)\\
&=(-\I)^{n-m}  \oint_{|z_1|=1}\oint_{|z_2|=1} \frac{\rmd z_1}{2\I \pi}\frac{\rmd z_2}{2\I \pi}e^{\frac{t}{2}(z_2-\frac{1}{z_1})} \sum_{k=a}^b \frac{e^{\frac{t}{2}(z_1-\frac{1}{z_2})}}{z_1^{k-n+1} z_2^{k-m+1}}\\
&=\frac{\I^{m-n} }{2}  \oint_{|z_1|=1}\oint_{|z_2|=1} \frac{\rmd z_1}{2\I \pi}\frac{\rmd z_2}{2\I \pi}e^{\frac{t}{2}(z_2-\frac{1}{z_1})} \frac{1}{z_1^{-n+1} z_2^{-m+1}} \left( \int_0^t \rmd t_1 e^{\frac{t_1}{2}(z_1-\frac{1}{z_2})} -\frac{2}{z_1-\frac{1}{z_2}}\right) \sum_{k=a}^b \left( \frac{1}{z_1^{k-1} z_2^{k}}-\frac{1}{z_1^{k} z_2^{k+1}}\right)
\end{split}
\end{equation}
where we can drop a term because it leads to an integrand without poles with respect to the complex variables $z_1$ and $z_2$, see \cite{Sasamoto_LargedeviationXX}, such that
\begin{equation}
\begin{split}
[\hat{J}^\dagger P_{\mathcal{S}}\hat{J}]_{n,m}&=\frac{\I^{m-n}}{2}   \oint_{|z_1|=1}\oint_{|z_2|=1} \frac{\rmd z_1}{2\I \pi}\frac{\rmd z_2}{2\I \pi}e^{\frac{t}{2}(z_2-\frac{1}{z_1})} \frac{1}{z_1^{-n+1} z_2^{-m+1}} \int_0^t \rmd t_1 e^{\frac{t_1}{2}(z_1-\frac{1}{z_2})} \left[\frac{1}{z_1^{a-1} z_2^{a}}-\frac{1}{z_1^{b} z_2^{b+1}}\right]\\
&=\frac{\I^{m-n}}{2}   \int_0^t \rmd t_1  \oint_{|z_1|=1}\oint_{|z_2|=1} \frac{\rmd z_1}{2\I \pi}\frac{\rmd z_2}{2\I \pi}e^{\frac{t}{2}(z_2-\frac{1}{z_1})} e^{\frac{t_1}{2}(z_1-\frac{1}{z_2})} \left[\frac{1}{z_1^{a-n} z_2^{a-m+1}}-\frac{1}{z_1^{b-n+1} z_2^{b-m+2}}\right]\\
\end{split}
\end{equation}
Upon rescaling $z_2 = \tilde z_2 \sqrt{\frac{t_1}{t}}$ and $z_1 = \tilde z_1 \sqrt{\frac{t}{t_1}}$, one obtains
\begin{equation}
\begin{split}
[\hat{J}^\dagger P_{\mathcal{S}}\hat{J}]_{n,m}=&\frac{\I^{m-n}}{2} \int_0^t \rmd t_1  \oint_{|\tilde z_1|=\sqrt{\frac{t_1}{t}}}\oint_{|\tilde z_2|=\sqrt{\frac{t}{t_1}}} \frac{\rmd \tilde z_1}{2\I \pi}\frac{\rmd \tilde z_2}{2\I \pi}\left(\sqrt{\frac{t}{t_1}}\right)^{n-m+1}   e^{\frac{\sqrt{t t_1}}{2}(\tilde z_1-\frac{1}{ \tilde z_1}+ \tilde z_2-\frac{1}{\tilde z_2})}\\& \left[\frac{1}{\tilde z_1^{a-n} \tilde z_2^{a-m+1}}-\frac{1}{\tilde z_1^{b-n+1} \tilde z_2^{b-m+2}}\right]\\
=&-\frac{\I^{n-m}}{2}   \int_0^t \rmd t_1  \left(\sqrt{\frac{t}{t_1}}\right)^{n-m+1} \left( J_{n-a+1}(\sqrt{tt_1})J_{m-a}(\sqrt{tt_1})-J_{n-b}(\sqrt{tt_1})J_{m-b-1}(\sqrt{tt_1}) \right)\\
\end{split}
\end{equation}
where we assumed that $ n,m\notin [a,b]$ such that $[J^\dagger P_{\mathcal{A}}J]_{n,m}(t=0)=0$. Let us now push $b \to \infty$ such that the corresponding boundary terms disappear. We also rescale $t_1=\tilde{t}/t$ such that
\begin{equation} 
[\hat{J}^\dagger P_{\mathcal{S}}\hat{J}]_{n,m} =-\frac{(\I t)^{n-m}}{2}   \int_0^{t^2} \rmd \tilde{t} \; \tilde{t}^{\frac{a-n-1}{2}}   J_{n-a+1}\left(\sqrt{\tilde{t}}\right)J_{m-a}\left(\sqrt{\tilde{t}}\right)\tilde{t}^{-\frac{a-m}{2}} 
\end{equation}

We see that we can rewrite $J^\dagger P_{\mathcal{S}}J=-\tilde{J}_1 \tilde{J}_2$ where $\tilde{J}_{1,2}$ are semi-discrete integral operators with kernels
\begin{equation}
\begin{split}
&\tilde{J}_1(n,\tilde{t})=\left(\frac{\I t }{\sqrt{\tilde{t}}}\right)^{n-a}  J_{n-a+1}\left(\sqrt{\tilde{t}}\right)\\
&\tilde{J}_2(\tilde{t},m)= \left(\frac{\I t}{\sqrt{\tilde{t}}}\right)^{a-m}  J_{m-a}\left(\sqrt{\tilde{t}}\right)\\
\end{split}
\end{equation}
where the measure of integration is $\frac{1}{2} \int_0^{t^2} \frac{\rmd \tilde{t}}{\sqrt{\tilde{t}}}$, on the interval $[0,t^2]$. We have thus established Eq.~\eqref{eq:KernelRelation}.

\subsection{Recovering the Moriya et al identity for the uncorrelated domain-wall}
\label{app:RecoverContinuousBessel}

In this subsection we show that, in the case of the uncorrelated domain-wall initial condition where $\hat{C}_0 = P_\mathbb{Z^-}$, the kernel $-\tilde{J}_2 \hat{C}_0 \tilde{J}_1$ in the Fredholm determinant of Eq.~\eqref{eq:ContinuousToDiscrete} coincides with the continuous Bessel kernel. The goal of the derivation is to show that our introduction of semi-discrete kernels is a way to recover the results of \cite{Sasamoto_LargedeviationXX} in the case of the uncorrelated domain-wall initial condition. More precisely, we wish to establish the following identity
\begin{equation}
\label{conjecture}
[-\tilde{J}_2 \hat{C}_0\tilde{J}_1](t_1,t_2) =  \left(\sqrt{\frac{t_2}{t_1}}\right)^{a}   \sqrt{t_1}  \ \ \frac{\sqrt{t_1} J_{a-1}'(\sqrt{t_1})   J_{a-1}(\sqrt{t_2}) - \sqrt{t_2}J_{a-1}(\sqrt{t_1})  J_{a-1}'(\sqrt{t_2}) }{t_1 - t_2} 
\end{equation}
where the domain-wall initial condition $\hat{C}_0 = P_\mathbb{Z^-}$ gives the left-hand side as
\be 
[-\tilde{J}_2 \hat{C}_0\tilde{J}_1](t_1,t_2) = - \sum_{k\in \mathbb{Z}^-} \left(\sqrt{\frac{t_2}{t_1}}\right)^{a-k}  J_{k-a}(\sqrt{t_1})  J_{k-a+1}(\sqrt{t_2}) \; .
\ee
Indexing the sum by the natural integers and inserting the contour integral representation of the Bessel function introduced in the previous section, we have
\begin{equation} 
\begin{split}
[-\tilde{J}_2 \hat{C}_0\tilde{J}_1](t_1,t_2) &= 
- \sum_{k\in \N}  \left(\sqrt{\frac{t_2}{t_1}}\right)^{a+k} J_{-k-a}(\sqrt{t_1})  J_{-k-a+1}(\sqrt{t_2}) \\
&= \oint_{|z_1|=1}\oint_{|z_2|=1} \frac{\rmd z_1}{2\I \pi} \frac{\rmd z_2}{2\I \pi} \sum_{k\in \N} \left(\sqrt{\frac{t_2}{t_1}}\right)^{a+k} \frac{1}{z_1^{a+k+1}  z_2^{a+k}  } e^{ \frac{\sqrt{t_1}}{2} (z_1 - \frac{1}{z_1} ) + \frac{\sqrt{t_2}}{2} (z_2 - \frac{1}{z_2} ) } 
\end{split}
\end{equation}
The change of variables $\tilde z_{1} = \frac{z_{1}}{\sqrt{t_{2}}},\tilde z_{2} = \sqrt{t_1}z_2  $ yields
\begin{equation} 
\begin{split}
&[-\tilde{J}_2 \hat{C}_0\tilde{J}_1](t_1,t_2) = \\
 & \frac{1}{\sqrt{t_1}}   \oint_{|\tilde z_1|=\frac{1}{\sqrt{t_2}}}\oint_{|\tilde z_2|=\sqrt{t_1}} \frac{\rmd \tilde z_1}{2\I \pi} \frac{\rmd \tilde z_2}{2\I \pi} \sum_{k\in \N}   \frac{1}{\tilde z_1^{a+k+1}  \tilde z_2^{a+k}  } e^{ \frac{\sqrt{t_1 t_2}}{2} (\tilde z_1 - \frac{1}{\tilde z_2}) - \sqrt{\frac{t_1}{t_2}} \frac{1}{2 \tilde z_1} + \sqrt{\frac{t_2}{t_1}}\frac{\tilde z_2}{2} } 
\\
&= \frac{1}{\sqrt{t_1}}   \oint_{|\tilde z_1|=\frac{1}{\sqrt{t_2}}}\oint_{|\tilde z_2|=\sqrt{t_1}}
\frac{\rmd \tilde z_1}{2\I \pi} \frac{\rmd \tilde z_2}{2\I \pi}  \frac{e^{- \sqrt{\frac{t_1}{t_2}} \frac{1}{2 \tilde z_1} + \sqrt{\frac{t_2}{t_1}}\frac{\tilde z_2}{2} }}{\tilde z_1 (\tilde z_1 \tilde z_2)^a} \sum_{k\in \N} \frac{1}{(\tilde z_1 \tilde z_2)^k} \left( \frac{1}{2} \int_0^{\sqrt{t_1 t_2}} \rmd \tilde{t} \ e^{\frac{\tilde{t}}{2} (\tilde z_1 - \frac{1}{\tilde z_2})} \left( \tilde z_1 -  \frac{1}{\tilde z_2 } \right) +1 \right) \\
&= \frac{1}{\sqrt{t_1}}   \oint_{|\tilde z_1|=\frac{1}{\sqrt{t_2}}}\oint_{|\tilde z_2|=\sqrt{t_1}}
\frac{\rmd \tilde z_1}{2\I \pi} \frac{\rmd \tilde z_2}{2\I \pi}  \frac{e^{- \sqrt{\frac{t_1}{t_2}} \frac{1}{2 \tilde z_1} + \sqrt{\frac{t_2}{t_1}}\frac{\tilde z_2}{2} }}{\tilde z_1 (\tilde z_1 \tilde z_2)^a} \sum_{k\in \N}   \frac{1}{2} \int_0^{\sqrt{t_1 t_2}} \rmd \tilde{t} \ e^{\frac{\tilde{t}}{2} (\tilde z_1 - \frac{1}{\tilde z_2})} \left( \frac{1}{\tilde z_1^{k-1} \tilde z_2^k} -  \frac{1}{\tilde z_1^k \tilde z_2^{k+1} } \right)  
\end{split} 
\end{equation}
where we have dropped the +1 term in the last equation, as it leads to a null contribution in the contour integrals. The telescopic sum simplifies such that

\begin{equation}
\begin{split}
[-\tilde{J}_2 \hat{C}_0\tilde{J}_1](t_1,t_2) &=\frac{1}{\sqrt{t_1}}   \oint_{|\tilde z_1|=\frac{1}{\sqrt{t_2}}}\oint_{|\tilde z_2|=\sqrt{t_1}}
\frac{\rmd \tilde z_1}{2\I \pi} \frac{\rmd \tilde z_2}{2\I \pi}  \frac{e^{- \sqrt{\frac{t_1}{t_2}} \frac{1}{2 \tilde z_1} + \sqrt{\frac{t_2}{t_1}}\frac{\tilde z_2}{2} }}{\tilde z_1 (\tilde z_1 \tilde z_2)^a} \frac{1}{2} \int_0^{\sqrt{t_1 t_2}} \rmd \tilde{t} \ e^{\frac{\tilde{t}}{2} (\tilde z_1 - \frac{1}{\tilde z_2})}   \tilde z_1  \\
&= \frac{1}{2 \sqrt{t_1}}  \int_0^{\sqrt{t_1 t_2}} \rmd \tilde{t}   \oint_{|\tilde z_1|=\frac{1}{\sqrt{t_2}}}\oint_{|\tilde z_2|=\sqrt{t_1}} \frac{\rmd \tilde z_1}{2\I \pi} \frac{\rmd \tilde z_2}{2\I \pi}  \frac{1}{(\tilde z_1 \tilde z_2)^a}  e^{\frac{\tilde{t}}{2} (\tilde z_1 - \frac{1}{\tilde z_2})}    e^{- \sqrt{\frac{t_1}{t_2}} \frac{1}{2 \tilde z_1} + \sqrt{\frac{t_2}{t_1}}\frac{\tilde z_2}{2}}
\end{split}  
\end{equation}
Scaling to $\tau = \frac{\tilde t}{\sqrt{t_1 t_2}}$, we have
\begin{equation}
[-\tilde{J}_2 \hat{C}_0\tilde{J}_1](t_1,t_2) = \frac{\sqrt{t_2}}{2 }  \int_0^{1} \rmd \tau   \oint_{|\tilde z_1|=\frac{1}{\sqrt{t_2}}}\oint_{|\tilde z_2|=\sqrt{t_1}} \frac{\rmd \tilde z_1}{2\I \pi} \frac{\rmd \tilde z_2}{2\I \pi}  \frac{1}{(\tilde z_1 \tilde z_2)^a}  e^{\frac{\tau \sqrt{t_1 t_2}}{2} (\tilde z_1 - \frac{1}{\tilde z_2})}    e^{- \sqrt{\frac{t_1}{t_2}} \frac{1}{2 \tilde z_1} + \sqrt{\frac{t_2}{t_1}}\frac{\tilde z_2}{2}}
\end{equation}
Changing variables again to $\dbtilde{z}_1 = \sqrt{t_2 \tau} \tilde z_1$ and $\dbtilde{z}_2 = \frac{1}{\sqrt{t_1 \tau}}\tilde z_2$, we obtain
\begin{equation}
[-\tilde{J}_2 \hat{C}_0\tilde{J}_1](t_1,t_2) = \frac{\sqrt{t_1}}{2 } \left(\sqrt{\frac{t_2}{t_1}}\right)^{a} \int_0^{1} \rmd \tau   \oint_{|\dbtilde{z}_1|=\sqrt{\tau}}\oint_{| \dbtilde{z}_2|=\frac{1}{\sqrt{\tau}}} \frac{\rmd \dbtilde{z}_1}{2\I \pi} \frac{\rmd \dbtilde{z}_2}{2\I \pi}  \frac{1}{(\dbtilde{z}_1 \dbtilde{z}_2)^a}  
e^{\frac{\sqrt{\tau t_1}}{2} ( \dbtilde{z}_1 - \frac{1}{\dbtilde{z}_1}) + \frac{\sqrt{\tau t_2}}{2} ( \dbtilde{z}_2 - \frac{1}{\dbtilde{z}_2}) }
\end{equation}
such that finally, evaluating the contour integrals to Bessel functions and introducing $\tilde{\tau} = \sqrt{\tau}$:
\begin{equation}
\begin{split}
[-\tilde{J}_2 \hat{C}_0\tilde{J}_1](t_1,t_2) = \frac{\sqrt{t_1}}{2 } \sqrt{\frac{t_2}{t_1}}^{a}
\int_0^{1}  
J_{a-1}(\sqrt{\tau t_1}) J_{a-1}(\sqrt{\tau t_2}) \rmd \tau 
=\sqrt{t_1} \left(\sqrt{\frac{t_2}{t_1}}\right)^{a}
\int_0^{1}  \tilde \tau  
J_{a-1}( \tilde \tau \sqrt{ t_1}) J_{a-1}(\tilde \tau\sqrt{ t_2})  \rmd \tilde \tau  
\end{split}
\end{equation}
This is readily computed from known integrals \cite{NIST:DLMF}, and simplified to 
\begin{equation}
\begin{split}
& [-\tilde{J}_2 \hat{C}_0\tilde{J}_1](t_1,t_2) =  \sqrt{t_1} \left(\sqrt{\frac{t_2}{t_1}}\right)^{a}
\frac{\sqrt{t_1}J_{a}(\sqrt{t_1}) J_{a-1}(\sqrt{t_2}) - \sqrt{t_2}J_{a-1}(\sqrt{t_1}) J_{a}(\sqrt{t_2})}{t_1 - t_2} \\
&= \sqrt{t_1} \left(\sqrt{\frac{t_2}{t_1}}\right)^{a}
\frac{\sqrt{t_1}\left( J_{a-1}'(\sqrt{t_1}) +\frac{a-1}{\sqrt{t_1}} J_{a-1}(\sqrt{t_1}) \right) J_{a-1}(\sqrt{t_2}) - \sqrt{t_2}J_{a-1}(\sqrt{t_1}) \left( J_{a-1}'(\sqrt{t_2}) +\frac{a-1}{\sqrt{t_2}} J_{a-1}(\sqrt{t_2}) \right)}{t_1 - t_2}  \\
&= \sqrt{t_1} \left(\sqrt{\frac{t_2}{t_1}}\right)^{a}
\frac{\sqrt{t_1} J_{a-1}'(\sqrt{t_1})   J_{a-1}(\sqrt{t_2}) - \sqrt{t_2}J_{a-1}(\sqrt{t_1})  J_{a-1}'(\sqrt{t_2}) }{t_1 - t_2} 
\end{split}
\end{equation} 
This concludes the derivation of Eq.~\eqref{conjecture}.

\section{Transport measure in the continuous setting}
\label{App:transportmeasure}

We prove here the formula giving the slope as a function of the Wigner function Eq.~\eqref{slopetransport}. Let $n(x,k,t)$ be a Wigner function so that its initial condition and evolution read

\be
n(x,k,t  =0)=n^{{\rm eq}}(x,k), \qquad n(x,k,t)  =n^{{\rm eq}}(x-kt,k)\, .
\ee
The density profile at any given position $x$ and time $t$ is obtained as
\begin{equation}
\rho(x,t)=\int_{x/t}^{\infty}\frac{\rmd k}{2\pi}n(x,k)
\end{equation}
and the total number of particles in the right half-space at a given
time
\begin{align}
N_{{\cal R}}(t) & =\int_{0}^{\infty}\rho(x,t)\rmd x\\
 & =\int_{0}^{\infty}\int_{x/t}^{\infty}n^{{\rm eq}}(x,k)\frac{\rmd k}{2\pi}\rmd x\\
 & =\int_{0}^{\infty}\int_{1}^{\infty}n^{{\rm eq}}(x,\frac{x}{t}k')\frac{x}{t}\frac{\rmd k'}{2\pi}\rmd x\\
 & =\int_{0}^{\infty}\int_{1}^{\infty}n^{{\rm eq}}(\frac{x}{t}k')\frac{x}{t}\frac{\rmd k'}{2\pi}\rmd x\\
 & =t\int_{0}^{\infty}\int_{1}^{\infty}n^{{\rm eq}}(x'k')x'\frac{\rmd k'}{2\pi}\rmd x'\\
 & =t\int_{0}^{\infty}\int_{1}^{\infty}n^{{\rm eq}}(X)\frac{X}{k'^{2}}\frac{\rmd k'}{2\pi}\rmd X\\
 & =t\int_{0}^{\infty}n^{{\rm eq}}(X)X\frac{1}{2\pi}\rmd X
\end{align}
where we made the serie of substitutions $k'=k\frac{t}{x}$ and $x'=x/t$, $X=x'k'$ 
We also suppose that the $x$ dependence of $n^{{\rm eq}}(x,k)$ is
always of the form $\Theta(-x)n(k)$. We thus see that the dependence
in time is linear no matter what and the slope is given by : 
\begin{equation}
\mu_{T}(n^{{\rm eq}}):=\int_{0}^{\infty}\frac{dk}{2\pi}kn^{{\rm eq}}(k)
\end{equation}
which has the simple interpretation of the sum over all propagating
modes indexed by $k$ weighted by their velocity $\frac{\rmd \epsilon(k)}{\rmd k}=k$.

\end{document}